\documentclass[letterpaper,11pt]{article}
\usepackage[margin=1in]{geometry}
\usepackage[utf8]{inputenc} 
\usepackage[T1]{fontenc}    
\usepackage[colorlinks=true,breaklinks=true,bookmarks=true,urlcolor=blue,citecolor=blue,linkcolor=blue,bookmarksopen=false,draft=false]{hyperref}
\usepackage{url}            
\usepackage{booktabs}       
\usepackage{multirow}
\usepackage{microtype}      
\usepackage{amsfonts,amsmath, amsthm,mathtools,stmaryrd}
\usepackage{xcolor}
\usepackage[inline]{enumitem}
\usepackage{appendix}
\usepackage{float, graphicx, subcaption}
\usepackage{bbold}
\usepackage{natbib}
\usepackage[mathscr]{euscript}
\usepackage{tikz}

\renewcommand{\cite}[1]{\citep{#1}}

\newtheorem{remark}{Remark}[section]

\newtheorem{proposition}{Proposition}[section]

\newtheorem{definition}{Definition}[section]

\def\argmin_#1{\underset{#1}{\mathrm{arg\,min\, }}}
\def\argmax_#1{\underset{#1}{\mathrm{arg\,max\, }}}

\makeatletter
\def\dasharrowfill@#1#2#3#4{%
        $\m@th
        \thickmuskip0mu
        \medmuskip\thickmuskip
        \thinmuskip\thickmuskip
        \relax
        #4#1\mkern2mu
        \xleaders\hbox{$#4\mkern2mu#2\mkern2mu$}\hfill
        \mkern2mu
        #3$%
}

\def\dashleftarrowfill@{\dasharrowfill@\leftarrow\relbar\relbar}
\def\dashrightarrowfill@{\dasharrowfill@\relbar\relbar\rightarrow}
\def\dashleftrightarrowfill@{\dasharrowfill@\leftarrow\relbar\rightarrow}
\def\dashLeftarrowfill@{\dasharrowfill@\Leftarrow\Relbar\Relbar}
\def\dashRightarrowfill@{\dasharrowfill@\Relbar\Relbar\Rightarrow}
\def\dashLeftrightarrowfill@{\dasharrowfill@\Leftarrow\Relbar\Rightarrow}

\providecommand*\xdashleftarrow[2][]{%
  \ext@arrow 0055{\dashleftarrowfill@}{#1}{#2}}
\providecommand*\xdashrightarrow[2][]{%
  \ext@arrow 0055{\dashrightarrowfill@}{#1}{#2}}
\providecommand*\xdashleftrightarrow[2][]{%
  \ext@arrow 0055{\dashleftrightarrowfill@}{#1}{#2}}
\providecommand*\xdashLeftarrow[2][]{%
  \ext@arrow 0055{\dashLeftarrowfill@}{#1}{#2}}
\providecommand*\xdashRightarrow[2][]{%
  \ext@arrow 0055{\dashRightarrowfill@}{#1}{#2}}
\providecommand*\xdashLeftrightarrow[2][]{%
  \ext@arrow 0055{\dashLeftrightarrowfill@}{#1}{#2}}
\makeatother
\usepackage{nicefrac}       
\usepackage{wrapfig}         
\usepackage{algorithm}
\usepackage{algpseudocode}
\usepackage{caption}
\usepackage{mwe}
\begin{document}
\title{An optimal transport based embedding to quantify the distance between playing styles in collective sports.}
\author{
Ali Baouan
\thanks{Centre de Math\'ematiques Appliqu\'ees, Ecole Polytechnique.}
\and 
Sergio Pulido
\thanks{Université Paris–Saclay, CNRS, ENSIIE, Univ Évry, LaMME.}
\and
Mathieu Rosenbaum
\thanks{Centre de Math\'ematiques Appliqu\'ees, Ecole Polytechnique.
}}

\date{}
\maketitle
\begin{abstract}
This study presents a quantitative framework to compare teams in collective sports with respect to their style of play. The style of play is characterized by the team's spatial distribution over a collection of frames. As a first step, we introduce an optimal transport-based embedding to map frames into Euclidean space, allowing for the efficient computation of a distance. Then, building on this frame-level analysis, we leverage quantization to establish a similarity metric between teams based on a collection of frames from their games. For illustration, we present an analysis of a collection of games from the 2021-2022 Ligue 1 season. We are able to retrieve relevant clusters of game situations and calculate the similarity matrix between teams in terms of style of play. Additionally, we demonstrate the strength of the embedding as a preprocessing tool for relevant prediction tasks. Likewise, we apply our framework to analyze the dynamics in the first half of the NBA season in 2015-2016.
\end{abstract}

\section{Introduction}

In collective sports such as football or basketball, performance relies on how players position themselves on the pitch. This is because teams have to advance the ball to score with minimal obstruction from the opposition. The nature of each sport gives rise to general principles of positioning and specific strategies that distinguish every team and period. For example, in football, a general principle is that it is commonly advised to spread in offense to force the opposition to cover more ground, while it is recommended to remain compact in defense, see \cite{moura2012quantitative}. Another example is that defensively, the last line of defense should be aligned vertically to optimize with respect to the offside rule. As for team specificities, one can cite positional play that has gained a lot of ground with the recent success of Barcelona, Manchester City, and the Spain national team. It is based on a rigid positioning mechanism where team shape is important and individual interpretation is minimal and constrained. This is in contrast to more flexible systems that allow more freedom and interpretation for players.
\\
\noindent
\\
In this work, we introduce a quantitative framework to compare the styles of play between two teams. The style of play is determined by the way the players distribute themselves over the pitch at each instant. Naturally, the shapes taken by a team capture a large part of their general strategy of play. In particular, a defensive team displays more low blocks, whereas an offensive team positions its players high up on the pitch. Thus, it can be useful to provide a metric that describes the similarity between playing styles based on the way the teams distribute over the pitch. To achieve this goal, we leverage tracking data to collect a series of frames from each team's games, where each frame represents a snapshot of the players' locations at a specific timestamp. Our approach is twofold: first, we define a distance metric at the frame level to assess the similarity between two given player configurations; then, we extend this metric to compare entire collections of frames, thereby capturing the overall style of play across multiple games. In \cite{tangclustering}, the authors propose using a deep representation derived from an auto-encoder to measure the distance between frames. This is inspired by the SoccerMap neural network which employs convolution layers to predict the pass success probability, see \cite{fernandez2021soccermap}. While it provides an efficient way to measure similarity between game situations, it does not allow for control over the representation generated by the hidden layers of the neural network. The loss function selected during the training of the neural network can affect the relative importance of various factors in the resulting embedding, but this influence is challenging to grasp. Therefore, it is difficult to precisely interpret the notion of similarity captured by the distance in this embedding space.
\\
\noindent
\\
 As an alternative, we choose to represent the frames as discrete probability measures and exploit the optimal transport framework to construct a distance between game situations. With our modeling, the sliced-Wasserstein metric gives us a notion of distance between two given frames. Optimal transport is relevant for the second step as well. In fact, it allows us to lift the distance between frames to compare teams with respect to collections of their frames.
\\
\noindent
\\
Optimal transport was initially introduced by Gaspard Monge in the form of the Monge problem, see \cite{monge1781memoire}. Given piles of sand and holes with known locations, the problem is determining where every unit of sand from the pile should be moved at minimal cost. Essentially, the Monge problem can be seen as an optimal matching problem, where we need to find a deterministic coupling between initial and final locations. Kantorovich later provided a relaxation that makes the problem more tractable. By interpreting the piles of sand and holes as probability distributions, the optimal transport determines the most efficient way to transport the mass of one probability distribution to match the second. The resulting optimal cost is referred to as the Wasserstein distance and provides a distance metric in the space of probability measures. This theory has significant applications across various fields, including economics, biology and finance, see \cite{santambrogio2015optimal}. In our context, optimal transport is used at two distinct levels. At the first level, the probability distributions under consideration are discrete and given by the players' locations on the pitch. At the second level, the probability distributions are the empirical distributions representing a collection of frames.
\\
\noindent
\\
To analyze the way a team occupies space at every timestamp, we view every frame as an empirical probability distribution where the atoms are the locations of the players in the team under consideration. This representation of data induces a permutation invariance between players. In this study, we consider that players are interchangeable and should play similar roles if they swap positions over a period of time. This approximation is justified as we want to analyze the dynamics of the shape of the team; and position changes that do not imply a formation change incur minimal difference. With this representation, the Wasserstein distance offers a tool to compare two frames that accounts for spatial proximity and that is interpretable. The intuition behind it is that two frames will be considered to be close to each other if one can optimally transport the players in the first frame to the locations in the second frame with minimal cost. The sliced-Wasserstein distance offers an alternative way to measure the proximity between probability distributions that is computationally more efficient, see \cite{bonneel2015sliced}. It is based on iteratively projecting the atoms of each frame along different directions, and summing the distances between the resulting one-dimensional distributions. It leverages the closed-form solution of the optimal transport problem in the one-dimensional case and allows us to establish an embedding of the frames in Euclidean space that preserves the distance.
\\
\noindent
\\
Through our methodology, we retrieve an embedding that captures the important features of the spatial distribution of players in a given frame. It is an embedding in Euclidean space that is well tailored for learning tasks. As an example, we consider a given classification task to demonstrate its usefulness as a preprocessing tool. We compare different input formats to predict the ball possession given players’ locations in a given frame. For this task, we compare models like logistic regression, multi-layer perceptrons (MLP), and convolutional neural networks (CNN). This additional layer of analysis illustrates that the proposed embedding and resulting distance efficiently capture meaningful signatures of teams' behaviors. 
\\
\noindent
\\
Given this distance between frames, the aim is to measure the similarity between two teams given collections of frames from their games. Collective sports such as football and basketball are games of moments. To compare the styles of play of two opposing teams during a game, it is not relevant to compare the positions they take chronologically timestamp by timestamp. A better approach is to match both teams' offensive and defensive moments when comparing. This is where the optimal transport framework can be applied. Having embedded the frames in a Euclidean space, and with a notion of distance, one can view collections of frames as the associated empirical probability distribution in the embedding space and consider the Wasserstein distance, see Section~\ref{sec:comparingcollections} for the definition of the similarity metric.
\\
\noindent
\\ 
Finally, as the number of frames in a team's season is large, computing the Wasserstein distance between two teams' collections of frames is costly. Therefore, we perform a quantization in the embedding space to retrieve probability distributions with a limited number of atoms, compressing information and reducing complexity. Additionally, this is convenient because the quantization of empirical measures through Llyod's algorithm is equivalent to K-means clustering. The retrieved quantization Voronoi regions can be interpreted and used as clusters of frames. As an example, we display and analyze the clusters retrieved from the collection of frames that appear in the games of a randomly selected Ligue 1 team. We then establish a similarity matrix between all Ligue 1 teams in the 2021-2022 season. We further exploit the methodology to derive similarity metrics during different phases of play and retrieve insightful conclusions. To evaluate the embedding's ability to capture key features of playing styles, we perform a team identity prediction task taking as input a collection of frames from a given team. We are able to evaluate the number of randomly selected and unseen samples needed to correctly recognize the style of play of a team. In this paper, we extend our scope beyond football teams and incorporate various datasets for a more comprehensive analysis. While our approach primarily focuses on the 2021--2022 Ligue 1 season, we also extend our scope to basketball in the Appendix to show further the relevance of our methodology.
\\
\noindent
\\
Our framework is particularly relevant for practitioners at two distinct levels. On the one hand, it can be used for coaching purposes. Specifically, the ability to measure distances between frames allows for the automatic detection of game situations that are similar to predefined scenarios. Additionally, the embedding provides a preprocessing tool that enables various learning algorithms to extract meaningful spatial features from frames. On the other hand, the framework offers a quantitative measure of proximity between collections of frames. This metric can be used to analyze an opponent's positional patterns before a game. It is also useful for scouting purposes as it enables us to investigate the playing style of the teams of prospective recruits and select the one with the suitable playing characteristics.
\\
\noindent
\\
The paper is organized as follows. In Section \ref{sec:backgroundmethodology}, we focus on building a distance between frames providing necessary background and demonstrating its application in the context of tracking data. We discuss the Wasserstein and sliced-Wasserstein distances and introduce the embedding of frames in Euclidean space. Furthermore, we evaluate the embedding through a possession prediction experiment, showcasing its effectiveness for frame-level analysis. In Section \ref{sec:comparingcollections}, we extend our framework to compare collections of frames. Here, we define quantization and its role in compressing information within large frame collections, enabling the establishment of a similarity metric between teams. We illustrate this approach by analyzing games from the 2021-2022 Ligue 1 season, including clustering of game situations and the construction of similarity matrices between teams. Additionally, we demonstrate the embedding's ability to capture playing style through a team identity prediction experiment. In the Appendix, we provide results of the same analysis applied to data from the first half of the 2015-2016 NBA season. We also include proofs of the embedding's injectivity and justifications for the chosen normalization of the similarity metric.

\section{An embedding for frame representation }
\label{sec:backgroundmethodology}
 This section introduces the concepts of optimal transport in the context of sports tracking data as well as the embedding we propose to represent spatial information at the frame level. It includes necessary definitions and
theoretical results that will serve as the foundation for our applications and analyses in Sections~\ref{sec:predictingpossession} and~\ref{section:results}.

\subsection{Tracking data}
The football dataset is provided by Stats Perform and lists $100$ games from the 2021-2022 Ligue 1 season with the trajectories of the players at a frequency of 25 frames per second. Additionally, for each frame a value is provided specifying which team has possession of the ball, when the possession is assigned. Each team has between 8 and 11 games where it is involved in the dataset. The list of used games can be found in Tables~\ref{tab:games_combined} in the Appendix. The basketball dataset is provided by SportVu and contains player and ball trajectories from 630 games in the 2015-2016 NBA season. For each game, a series of \textit{moments} records information about players, their location, the period and the game clock at a 25 frames per second frequency. Moments can overlap, in the sense that some timestamps can appear multiple times in the dataset.  We process them so that every frame is used once.  In both cases, we exclude all frames with less than $n$ players on the pitch as they do not fit our modeling, where $n=11$ for football and $n=5$ for basketball. Moreover, we rotate the pitch when necessary, with the convention that the team under consideration must attack on the right side. For the basketball dataset, the side that a team defends is determined by comparing the average relative positions of both teams during each half of the game. In the case of football, this is done by looking at the average team position at the first frame of the game, where each team is located in its designated side.
 \\
 \noindent
 \\
 For each game and for each team, we construct a sequence of location vectors $(x_{t,i})_{i\leq n}$ in $\mathbb{R}^{n\times 2}$. At each timestamp $t$, the $i^{\text{th}}$ location does not necessarily correspond to the same player, as there can be substitutions for example. Moreover, from one game to another, a player can be assigned to a different index $i=1,\dots,n$. In our framework, we look at each frame as a set of points, or equivalently as a discrete measure with the location of the players as atoms. As such, our approach is permutation invariant, and we consider the players to be equal in ability and effect if they exchange their location.
\\
\noindent
\\
The data points on which we aim to compare teams with respect to their spatial strategy are the frames, where we discard the locations of the opponent. To ensure invariance by permutation in our modeling, we transform these data points into the space of discrete uniform measures $ \mathcal{P}^{\text{u}}_{n}(\mathbb{R}^2)$ through the following map:
\begin{align*}
    \phi  \colon & \mathbb{R}^{n\times 2}  \longrightarrow  \mathcal{P}^{\text{u}}_{n}(\mathbb{R}^2) \\
             & (x_{i})_{\substack{i=1,\dots, n}}                         \longmapsto   \frac{1}{n}\sum\limits_{i=1}^{n}\delta_{x_{i}} .
\end{align*}
\noindent
This representation encodes the positional distribution of the players on the field regardless of their identities. 

\subsection{Wasserstein distance to compare frames}
\label{section:wasserstein}
\begin{figure}[t!]
    \centering
    \includegraphics[width=0.9\textwidth]{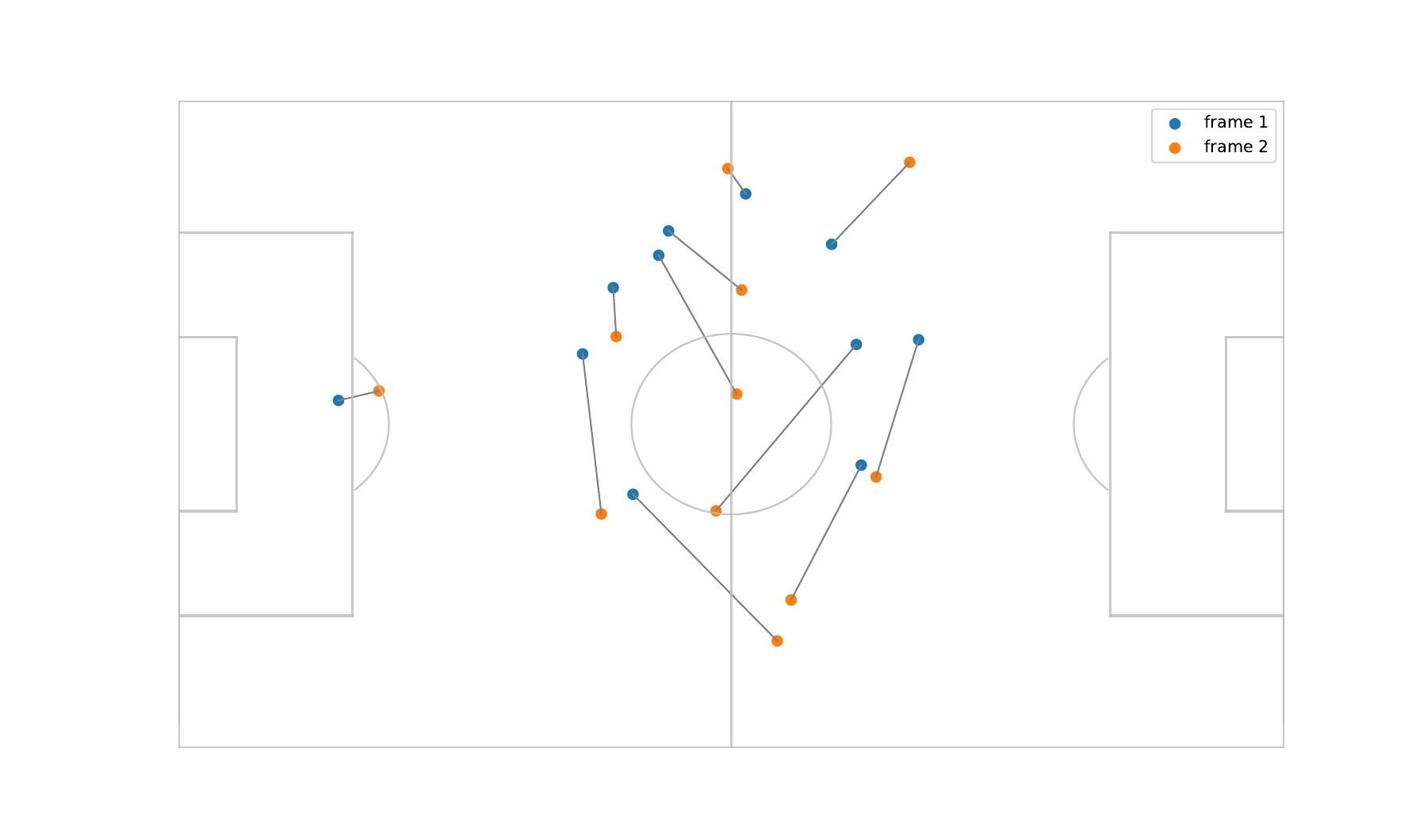}
    \vspace*{-.5em}
    \caption{The optimal one-to-one matching between two frames from a football game.}\label{fig:oneonematching}
\end{figure}

\noindent
With the representation of frames as discrete uniform measures, the Wasserstein distance is a natural metric to consider to compare frames. The Wasserstein distance is the solution of the optimal transport problem. Given a cost $c(x,y)$ of transport between two points $x$ and $y$ in space, the goal is to determine the best way to move mass distributed in space along some measure $\mu$ so that it becomes distributed along $\nu$ while minimizing the total cost of transport. Formally, the Wasserstein distance of order $p$ induced by a metric $d$ is defined as:
\begin{definition}
\label{def:wasserstein}
Let $(\mathcal{X}, d)$ be a Polish metric space, and let $\mu$ and $\nu$ be two probability measures on $\mathcal{X}$. A coupling of $\mu$ and $\nu$ is a probability measure $\pi$ on $\mathcal{X} \times \mathcal{X}$ such that for all measurable sets $A \subseteq \mathcal{X}$,
\[
\pi(A \times \mathcal{X}) = \mu(A) \quad \text{and} \quad \pi(\mathcal{X} \times A) = \nu(A).
\]
The collection of all such couplings is denoted by $\Pi(\mu, \nu)$. 
\noindent
The Wasserstein distance of order $p$ between $\mu$ and $\nu$ is defined as
\begin{align}
\label{eq:wasserstein}
W_p(\mu, \nu) = \left( \inf_{\pi \in \Pi(\mu, \nu)} \int_{\mathcal{X} \times \mathcal{X}} d(x, y)^p \, \pi(dx, dy) \right)^{1/p}.
\end{align}
\end{definition}
\noindent
\\
Here, the cost of transport between points $x$ and $y$ is given by $d(x,y)^p$. The optimal transport is determined by choosing an optimal coupling $\pi(dx,dy)$ that measures how much mass around $x$ should be transported to the location $y$. Naturally, the optimization is constrained as the total mass transported from $x$ should coincide with $\mu(dx)$ and the total mass transported to $y$ should match $\nu(dy)$.
\\
\noindent
\\
In many data-oriented applications such as this work, $\mu$ and $\nu$ are discrete distributions, and Equation~\eqref{eq:wasserstein} reduces to
\begin{equation}
\label{eq:discretewasserstein}
W_p^p(\mu,\nu)= \inf_{\substack{P \in \mathbb{R}_+^{n\times m}\\\sum\limits_{j=1}^mP_{i,j}=\mu_i \\ \sum\limits_{i=1}^nP_{i,j}=\nu_j}} \sum_{i=1}^n \sum_{j=1}^m P_{i,j} d(x_i,y_j)^p,
\end{equation}
where $\mu=\sum\limits_{i=1}^n \mu_i \delta_{x_i}$ and $\nu=\sum\limits_{i=1}^m \nu_i \delta_{y_i}$. In this setting, the mass is distributed along a finite number of locations $x_1,x_2,\dots,x_n$ following weights $\mu_1,\mu_2,\dots, \mu_n$ respectively. The mass is to be transported to new locations $y_1,y_2,\dots,y_m$ with a target distribution given by $\nu_1,\nu_2,\dots,\nu_m$. The coupling is now in matricial form and the optimal matrix $P$ in $\mathbb{R}^{n\times m}$ determines how the mass $\mu_i$ in each location $x_i$ is split over the locations $y_i$ of the second distribution. There is no closed-form solution to this optimization problem in general. However, the expression can be simplified in some special cases. When $\mu$ and $\nu$ have the same number of atoms $n=m$ and are uniform  $\mu_i=\nu_i=\frac{1}{n}$ for $i=1,\dots,n$, the expression can be reduced, using the Birkhoff-von Neumann theorem, to
\begin{equation}
\label{eq:permutationwasserstein}
W_p(\mu,\nu)=  \min_{\sigma \in S_n} \left (\frac{1}{n}\sum_{i=1}^n d(x_i,y_{\sigma(i)})^p\right )^{1/p}.
\end{equation}
\\
\noindent
In this case, the optimal transport problem is equivalent to an optimal matching problem. In particular, all the mass in any location of the first measure should be transported to one location of the second measure without splitting in the optimal case. This is interesting in the context of tracking data, since we choose to represent them as discrete uniform distribution with the same number of atoms. Essentially, the Wasserstein distance is interpreted as the minimal distance required to move the players from the first frame to fit the locations in the second. This is done by first determining the optimal one-to-one matching between players across the two frames and then summing the distances across coupled players. The solution of this optimization problem can be found using the Hungarian algorithm in $\mathcal{O}(n^3)$, see \cite{kuhn1955hungarian}. Figure \ref{fig:oneonematching} displays an example of the optimal correspondence to minimize the transport cost between two frames from a football game.
\\
\noindent
\\
 The Wasserstein distance provides an interpretable notion of distance between distributions. In this work, we use this optimal transport framework to quantify the distance between collections of points at two distinct levels. To do so, any collection of points in a metric space is viewed as its empirical distribution, \textit{i.e.}\ the uniform probability measure whose atoms are located at the points in the collection. 
 \\
 \noindent
 \\
 Here, we consider the Wasserstein distance to compare individual frames in the tracking data. We fall in the setting where the optimal transport between distributions is given by Equation~\eqref{eq:permutationwasserstein} because the frames have the same number of atoms. The Wasserstein distance provides a natural notion of distance between placement schemes over the pitch. In fact, two frames are close if the players in one can be shifted into the positions of the other with minimal total displacement. The Wasserstein distance determines the average distance the players in one frame should move to optimally occupy the locations in the second frame.
\\
\noindent
\\
The resulting metric between frames can be used to define the Wasserstein distance between collections of frames. These collections are treated as discrete empirical measures within the constructed metric space of frames, allowing Definition~\ref{def:wasserstein} to be applied. Section~\ref{sec:comparingcollections} provides further details on the similarity metric between collections of frames.

\subsection{The sliced-Wasserstein distance to compare frames}
The computation of the Wasserstein distance between discrete uniform measures with \( n \) atoms has a computational complexity of \( \mathcal{O}(n^3) \). While this polynomial complexity is manageable for comparing two frames with \( n = 5 \) or \( n = 11 \) in basketball and football, respectively, it can lead to prolonged computation times in the case of repetitive use as required in our framework in Section~\ref{section:quantization}. For example, there is no closed-form expression for the Wasserstein barycenter of a group of discrete measures, and its approximation requires iterative calculations of the Wasserstein distance.
\\
\noindent
\\
This issue can be mitigated by employing the sliced-Wasserstein distance. The sliced-Wasserstein distance is based on the observation that, in the special case where \( \mathcal{X} = \mathbb{R} \) and \( d \) is the Euclidean distance, the optimal transport between two empirical distributions of equal size, as defined in Equation~\eqref{eq:permutationwasserstein}, is explicitly determined by ordering the atoms
\begin{equation}
\label{eq:1Dwasserstein}
W_p(\mu,\nu)=  \left (\sum_{i=1}^n \frac{1}{n} |x_{(i)}-y_{(i)}|^p \right )^{1/p},
\end{equation}
where $x_{(1)}\leq x_{(2)}\leq \dots \leq  x_{(n)}$ and $y_{(1)}\leq y_{(2)}\leq \dots \leq  y_{(n)}$ are the order statistics. This significantly reduces the complexity of computing the distance between the probability measures as it suffices to sort the atoms.
\\
\noindent
\\
The sliced-Wasserstein distance is a metric derived from the Wasserstein distance to leverage the closed-form solution of the optimal transport problem in the one-dimensional case. Given probability measures in $\mathcal{X}=\mathbb{R}^d$, the idea is to sequentially project them onto one-dimensional axes and aggregate the distances between these different representations. The sliced-Wasserstein distance is formally defined as 
\begin{equation}
\label{eq:sliced}
    SW_p(\mu,\nu)= \left ( \int _{\mathbb{S}^{d-1}} W_p^p(\theta\# \mu, \theta \# \nu)\ d\theta \right )^{1/p},
\end{equation}
where $\mathbb{S}^{d-1}$ is the unit sphere in $\mathbb{R}^{d}$. Here, $\theta\# \mu$ is the push-forward of $\mu$ by the transformation  $\langle\theta,.\rangle$, or equivalently the image measure in $\mathbb{R}$ formed by projecting the mass of $\mu$ onto the axis determined by the direction of $\theta$. For example, if $\mu$ is an empirical measure $\mu=\frac{1}{n}\sum\limits_{i=1}^n \delta_{x_i}$, then $\theta\# \mu$ is also a uniform discrete measure in $\mathbb{R}$ located at the projections and given by $\frac{1}{n} \sum\limits_{i=1}^n \delta_{\langle\theta_l,x_{i}\rangle}$. Essentially, the sliced-Wasserstein distance is the sum of the Wasserstein distances between the empirical measures after projecting along the one-dimensional axes. It satisfies the metric axioms and is equivalent to the Wasserstein distance when considering distributions on a compact set, see \cite{bonnotte2013unidimensional}.
Finally, the expression in Equation (\ref{eq:sliced}) is intractable and is approximated in practice by 
\begin{equation}
    \label{eq:slicedgrid}
    \widehat{SW}_p(\mu,\nu,(\theta_{l})_{l\leq L})= \left ( \frac{1}{L} \sum_{l=1}^L  W_p^p(\theta_l \# \mu, \theta_l  \# \nu) \right )^{1/p},
\end{equation}
where the grid $\theta_1,\dots,\theta_L$ is usually drawn from the uniform distribution in $\mathbb{S}^{k-1}$. In this work, we aim at applying this distance to compare frames that correspond to uniform distributions with a fixed number of atoms. Thus, we choose a fixed grid $\theta_1,\dots,\theta_L$ that ensures that one can reconstruct the initial positions from the successive projections. This also guarantees that  $\widehat{SW}_p(\mu,\nu,(\theta_{l})_{l\leq L})$ verifies the axioms of a distance.

\subsection{An embedding of frames}
\label{sec:embedding}
\begin{figure}[t!]
    \centering
    \includegraphics[width=0.9\textwidth]{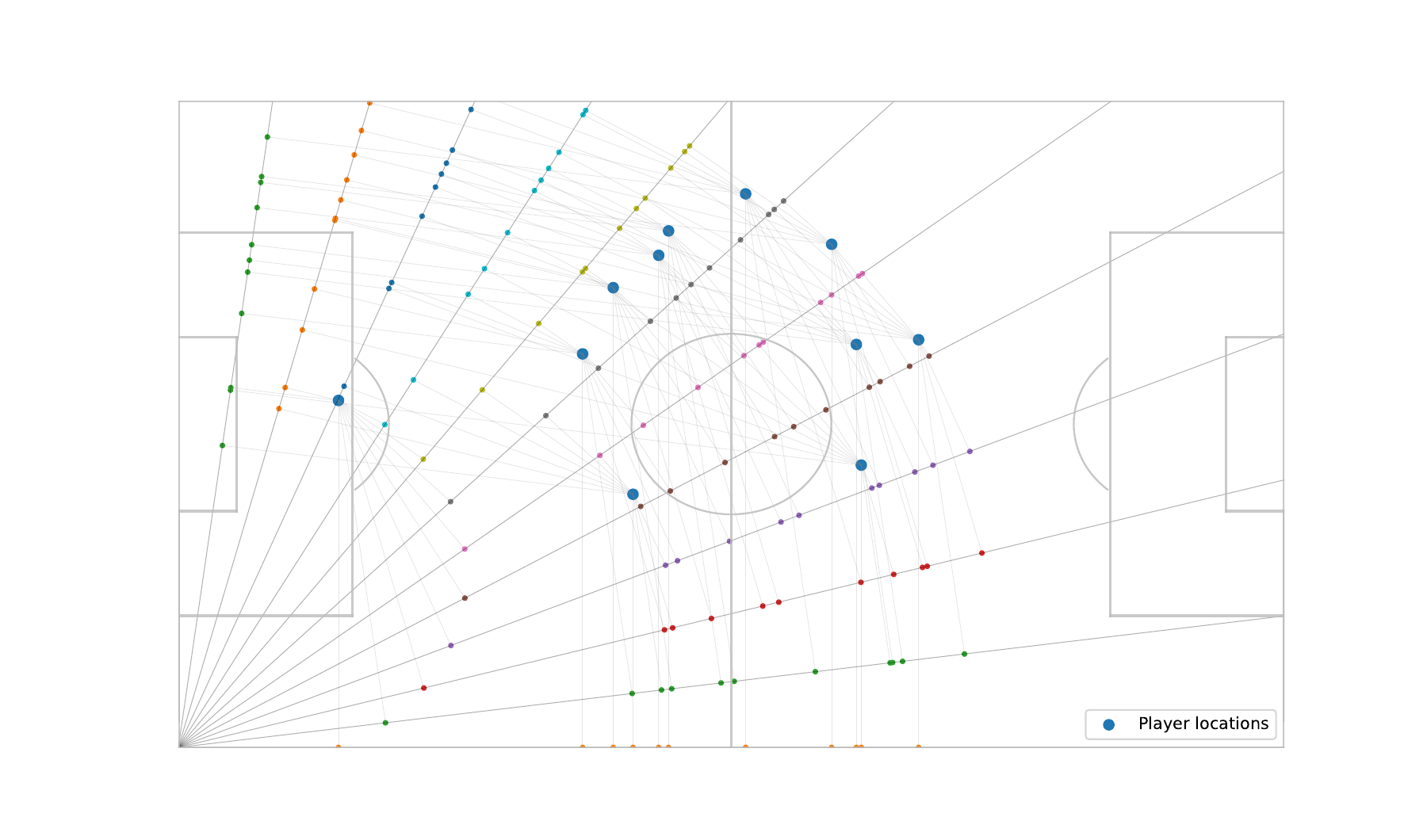}
    \vspace*{-.5em}
    \caption{Graphical representation of the embedding process of a frame.}\label{fig:exampleembedding}
\end{figure}

Given a grid of directions $\theta_1,\theta_2,\dots, \theta_L$ and two discrete uniform probability measures $\mu=\frac{1}{n}\sum_{i=1}^n \delta_{x_i}$ and $\nu=\frac{1}{n}\sum_{i=1}^n \delta_{y_i}$, the sliced-Wasserstein in Equation (\ref{eq:slicedgrid}) can be written as
\begin{equation*}
    \widehat{SW}_p(\mu,\nu,(\theta_{l})_{l\leq L})= \left ( \frac{1}{nL} \sum_{l=1}^L \sum_{i=1}^n \left | \langle\theta_l,x_{(i)}\rangle - \langle\theta_l,y_{(i)}\rangle \right |^p \right )^{1/p},
\end{equation*}
where $\langle\theta_l,x_{(1)}\rangle\leq \langle\theta_l,x_{(2)}\rangle\leq \dots \leq \langle\theta_l,x_{(n)}\rangle$ and $\langle\theta_l,y_{(1)}\rangle\leq \langle\theta_l,y_{(2)}\rangle\leq \dots \leq \langle\theta_l,y_{(n)}\rangle$ are the order statistics of the projections of the atoms $(x_i)_{i\leq n}$ and $(y_i)_{i\leq n}$ along the directions $\theta_l$ for $l=1,\dots,L$. Thanks to this property, we no longer need the Hungarian algorithm to find an optimal permutation matrix. The projection and sorting can be performed separately for each frame prior to calculating the distance. This is in contrast to the regular Wasserstein distance where the optimal matching has to be determined for each pair of frames. 
\\
\noindent
\\
The sliced-Wasserstein can be seen as an $\mathbb{L}^p$ distance between the vectors $\left ( \langle\theta_l,x_{(i)}\rangle \right )_{i\leq n,\  l\leq L} $ and $\left ( \langle\theta_l,y_{(i)}\rangle \right )_{i\leq n,\  l\leq L}$. This motivates the introduction of the following embedding of uniform probability measures on $\mathbb{R}^d$ with exactly $n$ atoms.
\begin{align*}
    \text{Proj}_{\theta}  \colon &  \mathcal{P}_n^u(\mathbb{R}^d)  \longrightarrow \mathbb{R}^{n\times L} \\
             & \mu = \frac{1}{n}\sum_{i=1}^n \delta_{x_i}                       \longmapsto   \left ( \langle\theta_l,x_{(i)}\rangle \right )_{i\leq n,\  l\leq L},
\end{align*}
where $\mathcal{P}_n^{\text{u}}(\mathbb{R}^d)$ is the set of uniform probability measures on $\mathbb{R}^d$ with exactly $n$ atoms. In the case of frames from tracking data, the grid of points along which we project is  
\begin{equation}
\label{eq:grid}
 \theta_l=\left (\cos(\frac{\pi (l-1)}{2L}),\sin(\frac{\pi (l-1)}{2L}) \right ),   
\end{equation}
for $l=1,\dots, L$, where $L=n+1$ and $n=11$ or $n=5$ for football and basketball respectively. Thus, given a vector of locations $x$ in $\mathbb{R}^{n\times 2}$, the final embedding is given by the vector $\text{Proj}_{\theta} \circ \phi (x)$ in $\mathbb{R}^{n\times L}$. Figure \ref{fig:exampleembedding} displays an example of the projections used for the embedding of a frame from a game of football.
\\
\noindent
\\
This grid satisfies the conditions to ensure injectivity with the following proposition:
\begin{proposition}
\label{lemma:distance}
Let $L\geq n+1$ and $\theta_1, \theta_2,\dots, \theta_L$ be unit vectors in $\mathbb{R}^d$ such that $\theta_i,\theta_j$ are non-collinear for all $i,j \leq L$.
Then $\widehat{SW}_p(.,.,(\theta_{l})_{l\leq L})$ is a distance in $\mathcal{P}^u_n(\mathbb{R}^d)$ and  $\text{Proj}_{\theta}$ is an injective distance-preserving map from $(\mathcal{P}^u_n(\mathbb{R}^d),\widehat{SW}_p(.,.,(\theta_{l})_{l\leq L}))$ to $(\mathbb{R}^{n\times L},\frac{1}{(nL)^{1/p}}\|.\|_p)$ .
\end{proposition}
\noindent
Using the grid in Equation~\eqref{eq:grid}, we obtain an embedding of uniform probability measures with $n$ atoms in Euclidean space, preserving the sliced-Wasserstein distance in Equation (\ref{eq:slicedgrid}). Proposition~\ref{lemma:distance} guarantees that the frame can be reconstructed from the embedding. However, it is not bijective as many points in $\mathbb{R}^{n\times L}$ do not correspond to the projections of any probability distribution. We use this embedding to encode the information about the locations of players in a given frame. This allows us to efficiently compute distances between frames and retrieve barycenters of a collection of frames. For the remainder of this paper, unless specified otherwise, player locations within a frame are represented in the embedding space. Moreover, we consider the Euclidean distance with $p=2$.

\subsection{Predicting possession with frame locations.}
\label{sec:predictingpossession}
To evaluate the quality of our embedding as a preprocessing tool, we consider a classification task aiming at predicting whether a team or its opponent holds possession, given only the players' locations. We discard all frames where possession of the ball is not assigned to any team, and merge the resulting tracking datasets of all teams. Furthermore, given the size of the dataset, we subsample it, keeping only one in every 10 frames to reduce computational cost. The resulting dataset comprises of $64,024$ frames and their target possession value. 
\\
\noindent
\\
The aim of this experiment is to assess the efficiency of the embedding as input for prediction tasks. To achieve this, we compare five different approaches to input representation:
\begin{itemize}
    \item \textbf{Tracking data}: Using the raw player positions.
    \item \textbf{Embedding}: Using our proposed embedding.
    \item \textbf{Image representation}: Encoding the pitch into a $10\times 10$ grid and counting the number of players in each cell.
    \item \textbf{Average player positions}: Using the mean \(x\) and \(y\) coordinates of all players in the frame.
    \item \textbf{Centered embedding}: Using the embedding after centering the locations in each frame by the mean location.
\end{itemize}
The last four approaches are permutation-invariant, offering robustness to changes in player identity. This is particularly important since we have merged data from different teams across different games. Therefore, they are naturally expected to perform better if we train a model on multiple team tracking data. We evaluate each feature construction using two classifiers:
\begin{itemize}
    \item \textbf{Logistic Regression}: A linear model used as a baseline for interpretability.
    \item \textbf{ Multi-layer perceptrons (MLP) and convolutional neural networks (CNN)}: Nonlinear models capable of capturing complex interactions in the data. The convolutional network is particularly useful for the image representation as it is capable of uniformly capturing local interactions between neighboring cells using a $3\times 3$ kernel.  
\end{itemize}
Table \ref{tab:possession_models} displays the out-of-sample accuracy using each model and each input features. For the logistic regression, we present the cross-validation accuracy while we choose display the out-of-sample accuracy after performing and train-test split for the neural networks. We observe the strong performance of our embedding across both logistic regression (81.66\%) and neural network models (82.26\%). This indicates that the embedding effectively captures positional and stylistic features relevant for predicting possession. The image representation also achieves competitive accuracy with  80.03\%  using the CNN, suggesting that spatial features encoded in grid-based formats are also meaningful. 
\begin{table}[ht]
    \centering
    \begin{tabular}{ccc}
        \toprule
        \toprule
        \textbf{Input Representation} & \textbf{Logistic Regression} & \textbf{MLP/CNN} \\
        \midrule
        Embedding & 81.66\% & \textbf{82.26}\% \\
        Tracking & 59.65\% & 76.77\% \\
        Image Representation & 73.71\% & 80.03\% \\
        Average Positions & 59.31\% & 59.92\% \\
        Centered Embedding & \textbf{81.81\%} & 80.63\% \\
        \bottomrule
        \bottomrule
    \end{tabular}
    \caption{Cross-validation accuracy of logistic regression and neural network models for possession prediction across input representations.}
    \label{tab:possession_models}
\end{table}
\\
\noindent
\\
In contrast, raw tracking data performs significantly worse. The lower accuracy of raw tracking (60.33\% with logistic regression) underscores the importance of preprocessing to extract meaningful patterns and to normalize the input between the teams and the games. Similarly, the performance of average positions (59.47\%) highlights the importance of the average position in the frame in determining the possession. Our embedding applied to centered frames is capable of delivering a strong accuracy despite filtering this information. This demonstrates that the embedding can encode useful information beyond the general area occupied on the pitch.
\\
\noindent
\\
These results demonstrate that permutation-invariant representations, such as our embedding and image-based grids, are well-suited for generalizing across teams. The embedding's robustness to player-specific variations makes it a particularly powerful tool for multi-team analyses, outperforming raw or oversimplified representations in identifying possession.
 Despite the inherent linearity of logistic regression, our embedding demonstrated strong classification performance.
\\
\noindent
\\
Having introduced a frame-level embedding in $\mathbb{R}^{n\times L}$  and demonstrated its utility for tasks such as possession prediction, we now extend this perspective to entire collections of frames. In the following section, we show how to compare two teams’ overall playing styles by measuring distances between collections of frames.
\section{Comparing collections of frames}
\label{sec:comparingcollections}
In Section~\ref{sec:backgroundmethodology}, we focused on the representation and analysis of individual frames. However, understanding a team’s style of play often requires examining its positioning patterns over an entire season or multiple matches. In this section, we formalize a notion of similarity between these large collections of frames. We first review the underlying definitions and introduce a quantization step to handle the computational challenges. Then, we present empirical results that illustrate how the analysis of collections of frames can reveal each team’s stylistic identity.
\subsection{Similarity between two collections of frames}
\label{section:quantization}
The embedding of frames in $\mathbb{R}^{n\times L}$ introduced in Section~\ref{sec:embedding} yields a distance between frames that preserves the sliced-Wasserstein distance. Building on this, a natural idea is to consider the empirical distribution of the collections of frames in the embedding space and then to use the Wasserstein distance of order 2 between the resulting distributions, as described in Equation \eqref{eq:wasserstein}. This procedure provides us with a notion of similarity between styles of play based on the spacial distribution during the season. Furthermore, it can be interpreted in the special case where the two collections have the same number of frames. The Wasserstein distance looks to match frames from the first collection to those in the second optimally so that the aggregated distance between frames is minimal. It is expected that two teams with similar styles of play generally occupy similar positions over a season, resulting in a small similarity metric. It should be noted that the matching may be made between frames happening at different times during the season.
\\
\noindent
\\
Formally, given two collections of frames $c_1=\{f_{1,1},f_{1,2},\dots,f_{1,N}\}$ and $c_2=\{f_{2,1},f_{2,2},\dots,f_{2,M}\}$ coming from the full season or a game of a team, the similarity between the two collections can be defined as 
\begin{equation}
    \label{eq:similarirtywasserstein}
    W_2\left (\frac{1}{N} \sum_{j=1}^N \delta_{f_{1,j}},\frac{1}{M} \sum_{j=1}^M \delta_{f_{2,j}}\right ),
\end{equation}
\noindent
where $f_{i,j}$ represent frames in the embedding space $\mathbb{R}^{n\times L}$. We recall that the Wasserstein distance requires a metric on the underlying space to be defined. In this section, it is the  distance between frames induced by the embedding and defined using optimal transport in Section~\ref{sec:embedding}. Consequently, the Wasserstein distance is applied at two levels, where the constructed metric on frames serves as the foundation for measuring distances between collections of frames.
\\
\noindent
\\
The number of frames in a team's season is very large, making the computation of the Wasserstein distance in Equation~\eqref{eq:similarirtywasserstein} impractical. The idea is, therefore, to consider compressed versions of each collection using quantization and then calculate the Wasserstein distance between the resulting distributions with $K$ atoms. Quantization aims at compressing information by selecting a finite set of representative data points and assigning them a weight proportional to the number of points they represent. This methodology also provides a framework to cluster the data points into groups that describe the phases of play that appear in the collections of frames of the team under consideration. 
\subsection{Quantization}
Let $Y$ be a $\mathbb{R}^d$-valued random variable with distribution $\mu$ such that $\mathbb{E}(\|Y \| ^2) <+\infty$. For $K \in \mathbb{N}$, the $K$-th quantization error for $\mu$ is given by, see \cite{graf2007foundations}:
\begin{equation}
   \label{eq:quantization}
       V_{K}(\mu)=\inf _{\substack{\alpha \subset \mathbb{R}^d \\  \mid \alpha \mid \leq K}} \mathbb{E} \left ( \min _{a \in \alpha} \|Y-a\|^2 \right ).
   \end{equation}
\noindent
For a set of representative points $\alpha$, we assign every point in space to the closest centroid $a$ in $\alpha$. The goal is to find the optimal subset $\alpha$ so that the expected dispersion induced by the assignments is minimal. When $\mu$ is the empirical distribution of some collection of data points, we recover the objective of the $K$-means clustering problem. In fact, every two data points are considered from the same cluster if they are assigned to the same centroid from $\alpha$. Therefore, optimal quantization allows us to cluster datasets simultaneously. In this work, the points we look to cluster are the frames in the embedding space, describing the positioning of a team. 
\\
\noindent
\\
Finally, we can also derive an alternative expression of the $K$-th quantization error
$$
V_{K}(\mu) =\inf _{\substack{\nu \in \mathcal{P}_K(\mathbb{R}^d) }}  W_2^2( \mu , \nu) ,
$$
\noindent
where $\mathcal{P}_K(\mathbb{R}^d)$ is the space of discrete probability measures with at most $K$ atoms. The minimizer of this problem is a projection onto the subspace of discrete measures with support less than $K$. In this form, optimal quantization provides an approximation of the measure $\mu$ with a limited number of atoms.
\\
\noindent
\\
Introduced in \cite{lloyd1982least}, Lloyd's algorithm is a fixed point algorithm to approximate the centroids. It is not guaranteed to converge as it could cycle in theory, but in practice, it generally converges to a stationary local minimum of Equation~\eqref{eq:grid}. Furthermore, the output is highly dependent on the initialization of the centroids. The \textit{kmeans++} algorithm provides a procedure to get initial centroids as spread out as possible, yielding a good starting quantization error, see \cite{kmeansplus}.
\\
\noindent
\\
In this work, we quantize distributions of frames embedded in $\mathbb{R}^{n\times L}$. Consequently, the distance in Equation~\eqref{eq:quantization} is given by the Euclidean norm within this embedding space, which is equivalent to the sliced-Wasserstein distance. As discussed in Section~\ref{sec:embedding}, this metric represents the optimal transportation between frames. 
\\
\noindent
\\
Given two collections of frames in the embedding space $\mu_1=\frac{1}{N} \sum\limits_{j=1}^N \delta_{f_{1,j}}$ and  $\mu_2=\frac{1}{M} \sum\limits_{j=1}^M \delta_{f_{2,j}}$, we consider $\hat{\mu}_1$ and $\hat{\mu}_2$  their respective $K$-quantizers found given by Lloyd's algorithm. For $t=1,2$, $\hat{\mu}_t$ is a discrete measure with $K$ atoms. These atoms correspond to the cluster centroids, with weights proportional to the number of frames assigned to each cluster. The quantization allows us to approximate the distance in Equation~\eqref{eq:similarirtywasserstein} with
\begin{equation*}
    \text{similarity}(c_1,c_2)=\sqrt{2}W_2\left (\hat{\mu}_1,\hat{\mu}_2 \right ).
\end{equation*}
Essentially, we first cluster the game situations for each team in the embedding space and then compare them with respect to their clusters and the weight of each cluster.

\begin{remark}
We scale the similarity metric by a factor of $\sqrt{2}$ to obtain a quantity that reflects the typical distance a single player travels. In fact, the similarity metric is based on the sliced-Wasserstein distance between frames, and this distance must be normalized to accurately represent the distance a player moves. This normalization is necessary because the sliced-Wasserstein distance averages one-dimensional distances across multiple projection axes, whereas the Euclidean distance in the plane sums the distances across the two axes. Proposition~\ref{prop:equivalence} provides bounds centered at $\frac{1}{\sqrt{2}}$, comparing the sliced-Wasserstein distance to the standard Wasserstein distance. After normalization, the similarity metric can be interpreted as the average distance a player needs to move from their position in the first set of frames to a position in the second set when optimally matched. A similarity score of zero indicates that the two teams have identical player positions, merely rearranged across the collection of frames.
\end{remark}
\noindent
It should be noted that we avoid directly using the Wasserstein distance for both of our use cases for different reasons:
\begin{itemize}
   \item \textbf{Comparing collection of frames:} In this case, the large number of atoms encourages us to first use to quantization to reduce the dimensionality before computing the Wasserstein distance. 
\item \textbf{Comparing frames:} Although the number of atoms is low, the repetitive use of Wasserstein distance encourages the adoption of an alternative metric. In particular, Lloyd's algorithm requires the computation of the barycenter to determine the optimal quantization. The approximation of the Wasserstein barycenter is performed using an iterative procedure that requires repeated calculation of the Wasserstein distance between all the frames and the barycenter, see \cite{cuturi2014fast}. In \cite{domazakis2020clustering}, the authors propose the use of K-means clustering in the space of probability measures using the Wasserstein distance and the Wasserstein barycenter despite this difficulty. However, the non-flat geometry induced on the space of probability measures $\mathcal{P}(\mathbb{R}^2)$ by the $W_2$ less suitable for centroid-based algorithms such as Lloyd, see \cite{zhuang2022wasserstein}. In particular, a set of irregularities can be observed when considering the properties of the barycenter compared to Euclidean space. For example, for $x_1,\dots,x_l$ in $\mathbb{R}^2$, the barycenter $\overline{x}$ belongs to the convexhull of $\{x_1,\dots,x_l\}$ but this property does not hold for probability measures. They overcome this disadvantage of the Wasserstein metric by considering an alternative objective function for the K-means clustering. Since our goal is not only to determine the clusters but also to use quantization as an approximation of a collection of frames, we choose to leverage the sliced-Wasserstein following the idea in \cite{luan2023automated}. In this work, the authors introduce a K-means algorithm to cluster probability measures based on the sliced-Wasserstein distance where the centroids are updated in the projection space. The quantization step using Lloyd's algorithm in the embedding space is equivalent to the algorithm they introduce. Given the specific nature of the probability measures associated with frames, with a fixed and limited number of atoms, the choice of a fixed grid of projection axis is justified and provides us with an injective embedding that preserves the sliced-Wasserstein distance. Additionally, It provides us with a preprocessing tool to create inputs that are sensitive to spatial proximity for machine learning models. It should be noted that the centroids that are computed in the embedding space  during Lloyd's algorithm do not necessarily correspond to an existing frame as the encoding is not surjective. 

\end{itemize}

\subsection{Results and applications}
\label{section:results}

In this section, we present results of the optimal transport framework to analyze the spatial distribution of players on the field in football. First, we display an example of clusters retrieved from the quantization process when applied to a collection of frames. Then, we show the resulting similarity matrix between Ligue 1 teams. Finally, we evaluate the ability of the embedding in capturing a team's playing style through team identity prediction experiment.

\subsubsection{Clustering of frames}
\begin{figure}[t!]
    \centering
    \begin{minipage}{0.9\textwidth}
        \centering
        \scalebox{0.4}{\includegraphics{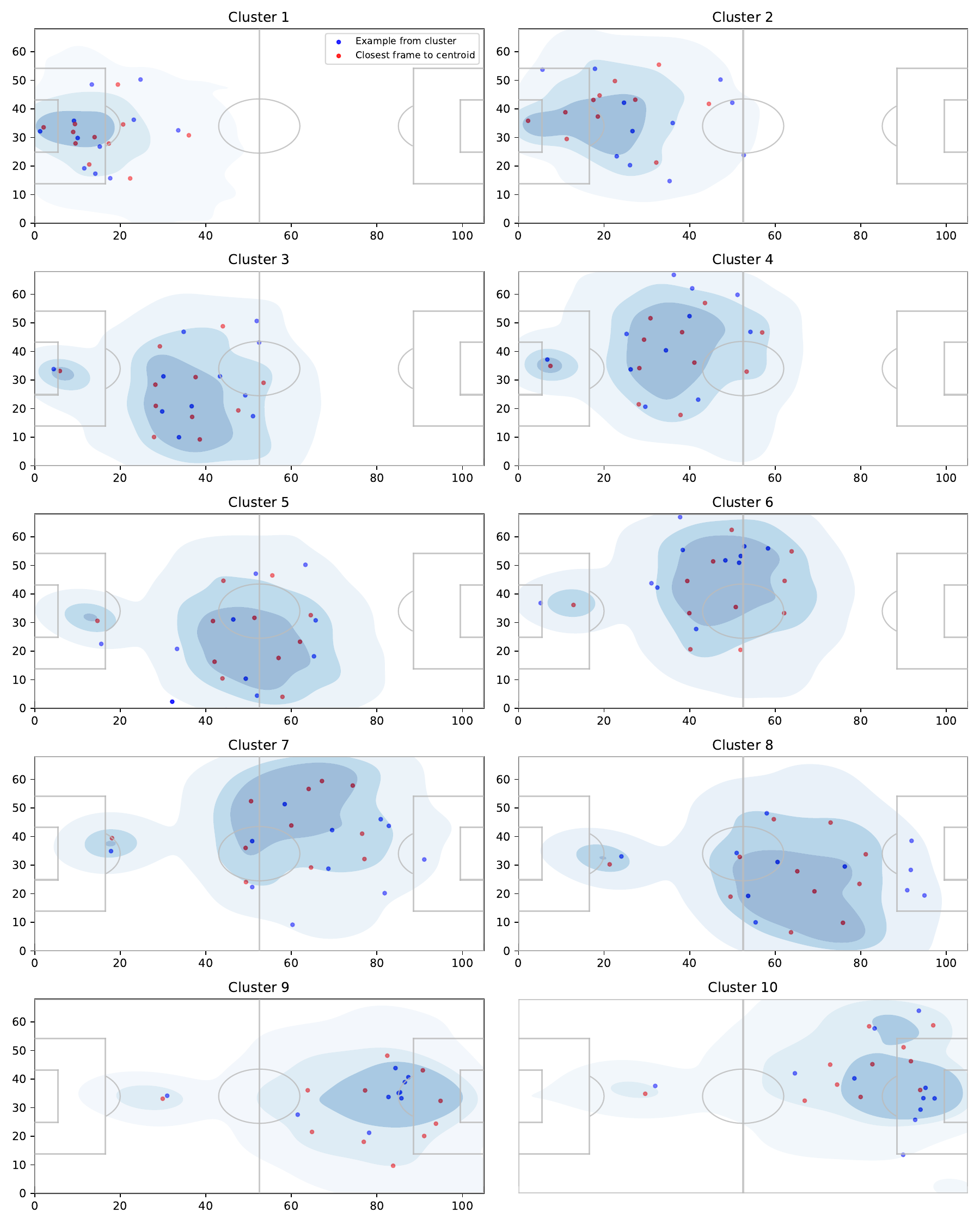}}\vspace*{-.5em}
        \captionof{figure}{Example of $10$ clusters observed in the frames of Brest during its games. For each cluster, the closest frame to the barycenter and a random example from the cluster are shown. The shaded blue color represents the total density of player locations taken from a random sample of 100 frames from each cluster.}\label{fig:clusteringrennesbrest}      
    \end{minipage}
\end{figure}

We apply Lloyd's algorithm introduced in Section \ref{section:quantization} to quantize the collection of frames in the games of Brest, a randomly selected Ligue 1 team, to retrieve $10$ clusters of frames.  The clustering algorithm is performed over the embedding space constructed with the grid in Equation \eqref{eq:grid} for $L=12$.
\\
\noindent
\\
Figure \ref{fig:clusteringrennesbrest} shows the resulting clusters. For each cluster, we display the closest frame to the centroid and a random frame sampled from the cluster. Table \ref{tab:cluster_possession} displays the frequency of occurrence of every cluster during the game and the average ball possession of Brest in the frames in the cluster. We observe that the clusters distinguish different phases of play. For instance, we can recognize in cluster 1 and 2 the frames where the team is organized as a deep low block, either skewed to the right or the left respectively. Clusters 3 and 4 represent game states where the team is organised as a mid block. In particular, the team has an average possession of $33.08\%$ and $40.19\%$ respectively in these clusters. The fact that these clusters are used more than the first two can be explained by the fact that they can arise both in defensive and offensive sequences, as most teams resort to build-up from the back to construct their attack.  Clusters 5 and 6 capture situations with the team organized in the middle of the pitch but slightly more advanced than the prior clusters. Finally, the last four clusters represent more advanced dispositions, and this is reflected in the average values of possession observed. 

\begin{table}[ht]
\centering
\begin{tabular}{l c c}
\toprule
\toprule
\textbf{Cluster} & \textbf{Percentage of Frames} & \textbf{Average Possession} \\
\midrule
 1  & 9.73\%  & 22.34\% \\
 2  & 8.26\%  & 27.96\% \\
 3  & 11.10\% & 33.08\% \\
\midrule
 4  & 13.30\% & 40.19\% \\
 5  & 11.86\% & 41.88\% \\
 6  & 12.31\% & 48.80\% \\
\midrule
 7  & 9.51\%  & 61.02\% \\
 8  & 9.72\%  & 63.14\% \\
 9  & 7.93\%  & 75.47\% \\
 10 & 6.28\%  & 76.52\% \\
\bottomrule
\bottomrule
\end{tabular}
\caption{Distribution of percentage of frames and average possession across clusters.}
\label{tab:cluster_possession}
\end{table}

\subsubsection{Similarity metric between frame collections}

\begin{figure}[t!]
    \centering
    \begin{minipage}{0.9\textwidth}
        \centering
        \scalebox{0.46}{\includegraphics{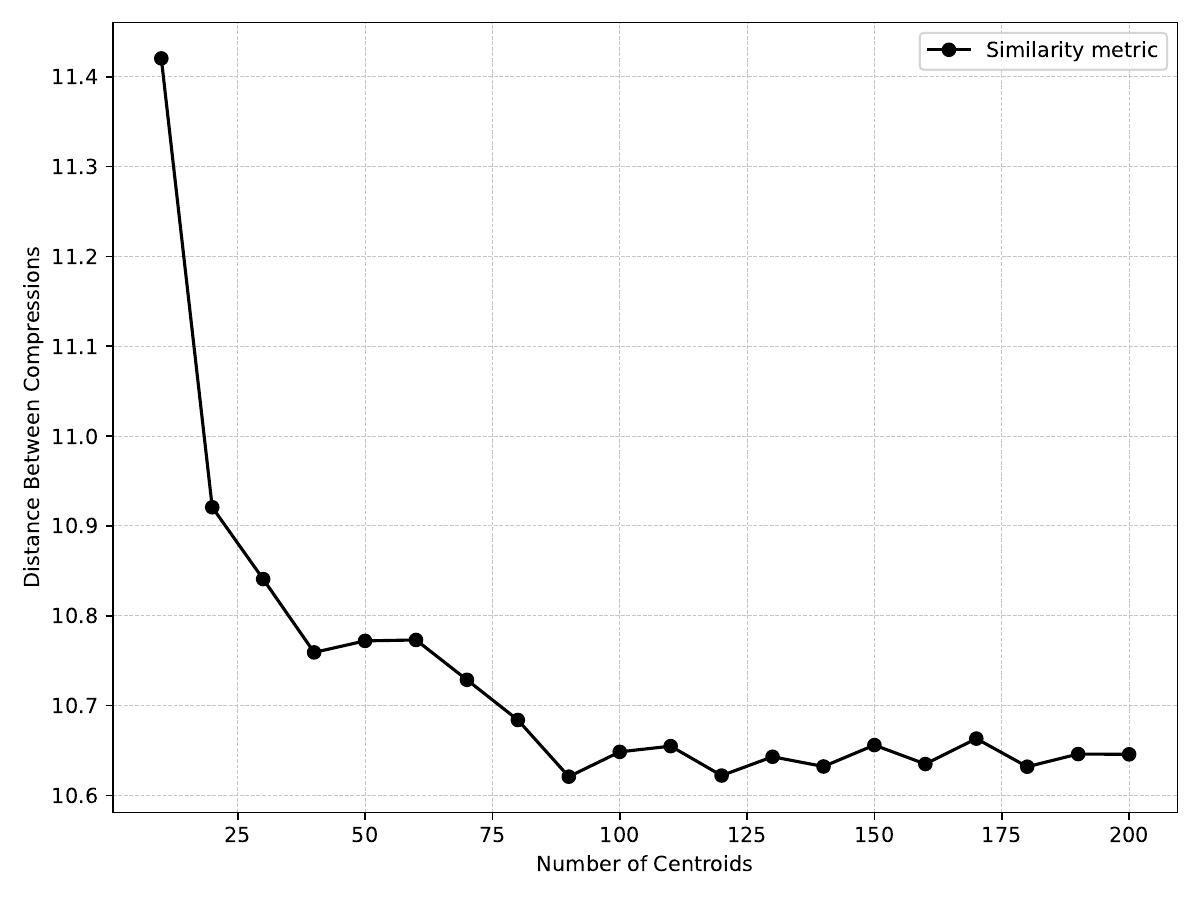}}\vspace*{-.5em}   
    \end{minipage}
    \caption{The evolution of the distance between Metz and PSG as a function of the number of centroids considered.}
    \label{fig:evolutionncentroids}
\end{figure}

\noindent
In this section, we estimate the similarity between the playing styles of different teams based on the collection of frames observed in their games. We consider the empirical distribution in $\mathbb{R}^{11\times 12}$ of the embedded frames from the games of each team. Frames with fewer than $11$ player location data points are excluded, as these instances often correspond to situations where a team has received a red card.
\\
\noindent
\\
For each collection of frames, we quantize their empirical distribution in the embedding space using $100$ centroids. This choice ensures that the quantized distribution is a close approximation of the original distribution for each team's frames. Figure~\ref{fig:evolutionncentroids} illustrates the evolution of the distance between Metz and PSG, two randomly selected Ligue 1 teams, as a function of the number of centroids. Beyond $100$ centroids, the estimated distance stabilizes, confirming the adequacy of this choice.

\noindent
\begin{figure}[t]
    \centering
    \includegraphics[width=0.8\textwidth]{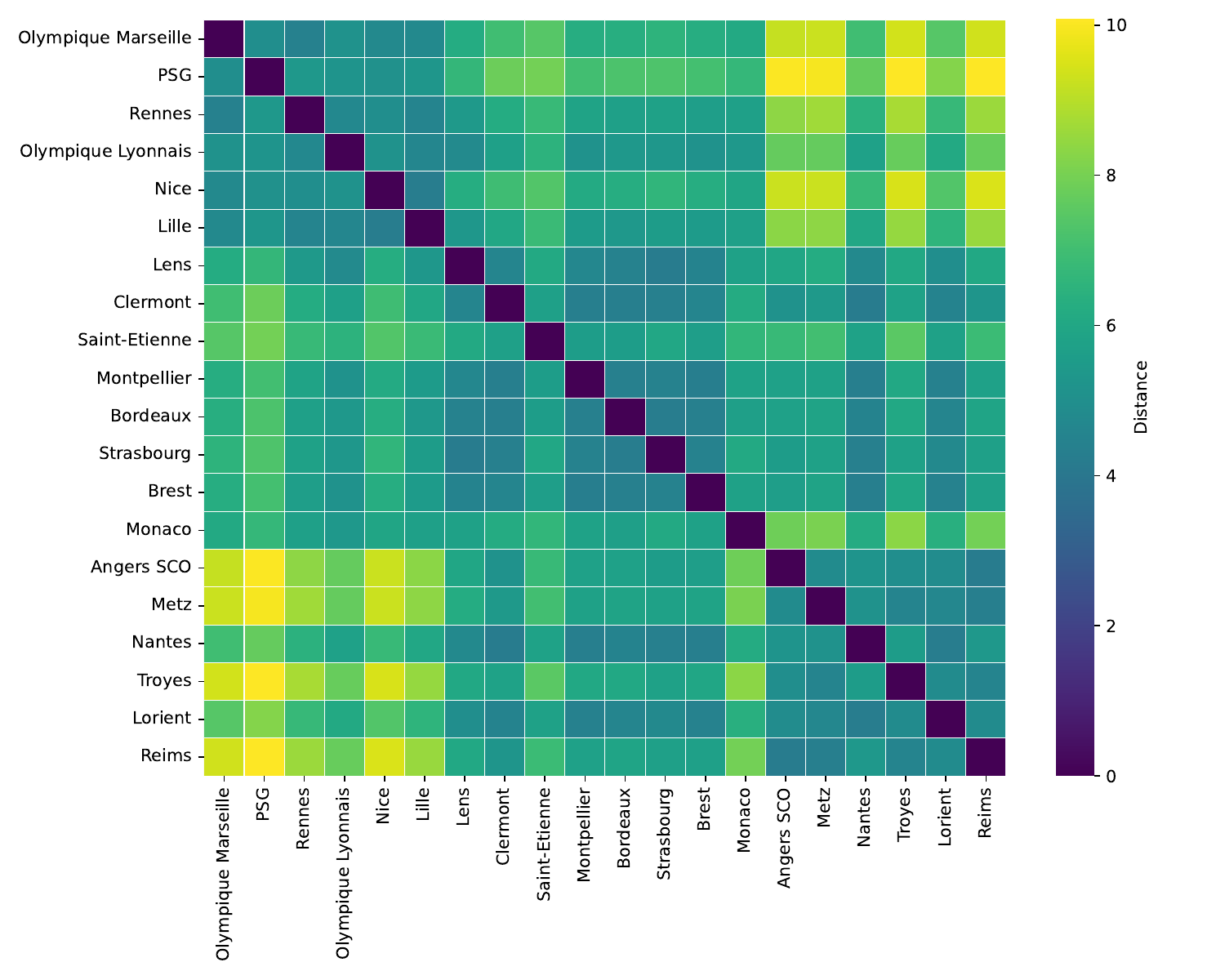}
    \caption{Similarity matrix between the collections of frames from the games of each team. Rows and columns are sorted by possession percentage.}
    \label{fig:similaritiesrennes}
\end{figure}

\begin{figure}[t]
    \centering
    \includegraphics[width=0.8\textwidth]{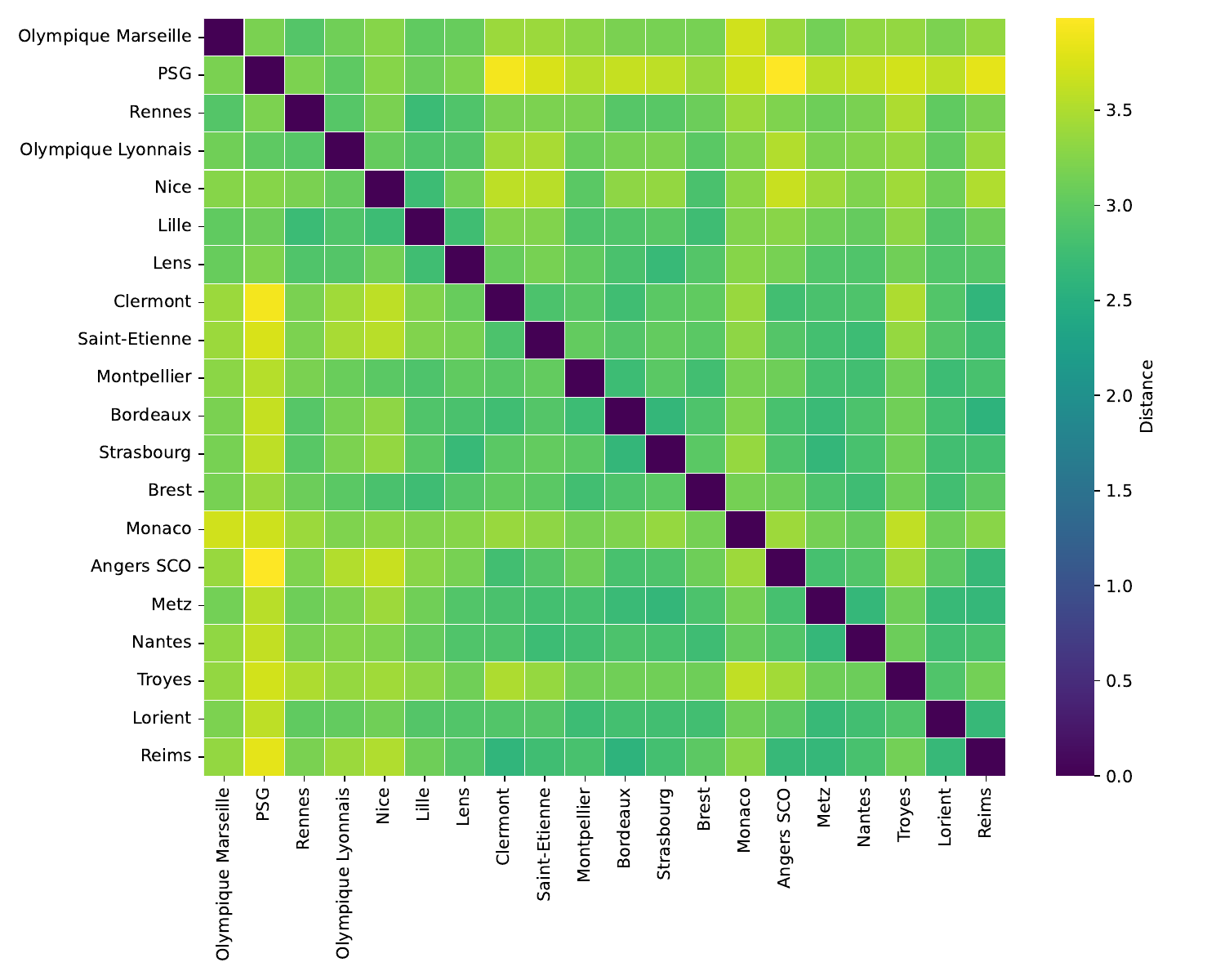}
    \caption{Similarity matrix between the collections of frames from the games of each team after centering the data.}
    \label{fig:similaritiesrennescentered}
\end{figure}

\\
\noindent
\\
Figure~\ref{fig:similaritiesrennes} displays the distances between the quantized distributions of frames for each team. The rows and columns are sorted in ascending order based on the possession percentage of the corresponding teams in the analyzed games. A complete list of teams, ranked by possession values, is provided in Table~\ref{tab:team_metrics_percentages} in the Appendix. Additionally, Table~\ref{tab:sum_distances_team} lists the sum of distances to the other teams and is sorted in descending order.
\\
\noindent
\\
The distance matrix allows us to identify the team displaying the highest degree of similarity to a given one in terms of positional patterns. Large distance values tend to appear far from the diagonal, which is expected given the sorting by possession. Generally, teams with higher possession occupy advanced areas of the pitch, while lower possession corresponds to deeper positioning. To support this observation, we measure the average possession for each team across all its frames. Next, for every pair of teams, we compute the absolute difference between their average possession values. We find a correlation of 66.49\% between these possession differences and our similarity metric derived from the distance matrix.
\\
\noindent
\\
To account for potential bias caused by the areas of the pitch occupied during possession, we recalculate the similarity matrix after centering the player locations in each frame by subtracting the average location in the frame. This adjustment focuses on relative player positions rather than absolute locations. Figure~\ref{fig:similaritiesrennescentered} shows the updated similarity matrix, where the effect persists, with a correlation of $54.72\%$ with the absolute difference of possession. This means that teams with similar possession values will generally display similar relative positions between players in their frames.
\begin{table}[ht]
\centering
\begin{tabular}{l c}
\toprule
\toprule
\textbf{Team} & \textbf{Value} \\
\midrule
PSG                     & 139.34 \\
Troyes                  & 129.67 \\
Olympique Marseille     & 127.45 \\
\midrule
Nice                    & 127.32 \\
Metz                    & 126.43 \\
Reims                   & 125.91 \\
\midrule
Angers SCO              & 125.05 \\
Saint-Etienne           & 123.58 \\
Monaco                  & 121.79 \\
\midrule
Rennes                  & 118.39 \\
Lille                   & 115.30 \\
Olympique Lyonnais      & 110.59 \\
\midrule
Lorient                 & 105.40 \\
Clermont                & 103.86 \\
Nantes                  & 103.33 \\
\midrule
Lens                    & 102.68 \\
Strasbourg              & 102.53 \\
Bordeaux                & 101.76 \\
\midrule
Brest                   & 101.18 \\
Montpellier             & 101.13 \\
\bottomrule
\bottomrule
\end{tabular}
\caption{Sum of distances to other teams.}
\label{tab:sum_distances_team}
\end{table}
\\
\noindent
\\
From Table~\ref{tab:sum_distances_team}, we observe that PSG has the highest average distance compared to other teams. The largest distance, measured at 10.07, is between PSG and Troyes. This indicates that, on average, a player from Troyes needs to move 10.07 meters to match a position in a PSG frame. This significant distance is largely explained by PSG's tendency to adopt more advanced positions on the pitch. In contrast, PSG shows smaller distances to other possession-based teams, such as Olympique Marseille. After centering the frames, Olympique Lyonnais emerges as the team most similar in style to PSG, as centering emphasizes the relative shapes of player formations rather than their absolute positions. The smallest distance is observed between Reims and Angers SCO at 4.18, reflecting their similar positional styles. This similarity remains evident even after centering the frames, where their distance reduces to 2.67, further highlighting their shared spatial patterns.
\\
\noindent
\\
This framework provides a method to measure the similarity between positional schemes. The results align with possession values and capture specificity that goes beyond the average location of the team. This methodology can be extended to compare playing styles across any collection of frames. In particular, Table~\ref{tab:distance_comparison_centered} displays, for each team, the distance between the collection of frames when in possession and out of possession. The table on the right shows the distance when the frames are centered by subtracting the mean location prior to the similarity metric calculation. The values are sorted in ascending order based on the distance in each table. This analysis provides a quantitative metric of how a team changes strategy with and without the ball. For example, we observe that PSG shows the second lowest distance. This is not surprising, as PSG consistently occupies advanced positions on the pitch both in attack and in defense through pressing. However, it is ranked 14th in terms of distance after centering, which suggests a change in shape when out of possession. In comparison, Montpellier shows small distances both with and without centering, indicating that their formation tends to be preserved in defense. In contrast, teams like Nantes display large distance values, suggesting significant changes in strategy between attack and defense. Finally, the large distance observed for Brest aligns with the clusters shown in Figure~\ref{fig:clusteringrennesbrest}, where we observe both low blocks and high lines.
\begin{table}[ht]
    \centering
    \begin{subtable}[t]{0.35\textwidth}
        \centering
        \begin{tabular}{lr}
        \toprule
        \toprule
        \textbf{Team} & \textbf{Distance} \\
        \midrule
        Montpellier & 6.73 \\
        PSG & 6.80 \\
        Reims & 6.91 \\
        \midrule
        Olympique Lyonnais & 6.97 \\
        Bordeaux & 7.36 \\
        Monaco & 7.39 \\
        \midrule
        Clermont & 7.57 \\
        Strasbourg & 7.57 \\
        Lens & 7.59 \\
        \midrule
        Rennes & 7.68 \\
        Olympique Marseille & 7.69 \\
        Lille & 7.77 \\
        \midrule
        Metz & 7.79 \\
        Nice & 7.88 \\
        Troyes & 7.89 \\
        \midrule
        Brest & 7.98 \\
        Angers SCO & 7.99 \\
        Lorient & 8.03 \\
        \midrule
        Saint-Etienne & 8.06 \\
        Nantes & 8.94 \\
        \bottomrule
        \bottomrule
    \end{tabular}
    \end{subtable}
    \hfill
    \begin{subtable}[t]{0.6\textwidth}
        \centering
        \begin{tabular}{lr}
        \toprule
        \toprule
        \textbf{Team} & \textbf{Distance after Centering} \\
        \midrule
        Montpellier & 3.62 \\
        Reims & 3.75 \\
        Clermont & 3.77 \\
        \midrule
        Olympique Lyonnais & 3.78 \\
        Angers SCO & 3.94 \\
        Lens & 3.96 \\
        \midrule
        Saint-Etienne & 3.96 \\
        Nice & 4.00 \\
        Bordeaux & 4.01 \\
        \midrule
        Brest & 4.01 \\
        Lille & 4.03 \\
        Lorient & 4.05 \\
        \midrule
        Olympique Marseille & 4.09 \\
        PSG & 4.10 \\
        Metz & 4.10 \\
        \midrule
        Monaco & 4.12 \\
        Rennes & 4.17 \\
        Troyes & 4.17 \\
        \midrule
        Strasbourg & 4.21 \\
        Nantes & 4.24 \\
        \bottomrule
        \bottomrule
    \end{tabular}
    \end{subtable}
    \caption{Comparison of team distances before and after centering frame locations. \textbf{Left}: Distance between frames in possession and out of possession for each team using the embedding of raw locations without centering. \textbf{Right}: Distance between frames in possession and out of possession for each team using the embedding of centered frame locations. The teams are sorted in ascending order of distance.}
    \label{tab:distance_comparison_centered}
\end{table}

\subsubsection{Predicting team identity using collections of frames}
\label{sec:teamidentity}
\begin{figure}[t!]
    \centering
    \resizebox{\textwidth}{!}{%
    \includegraphics{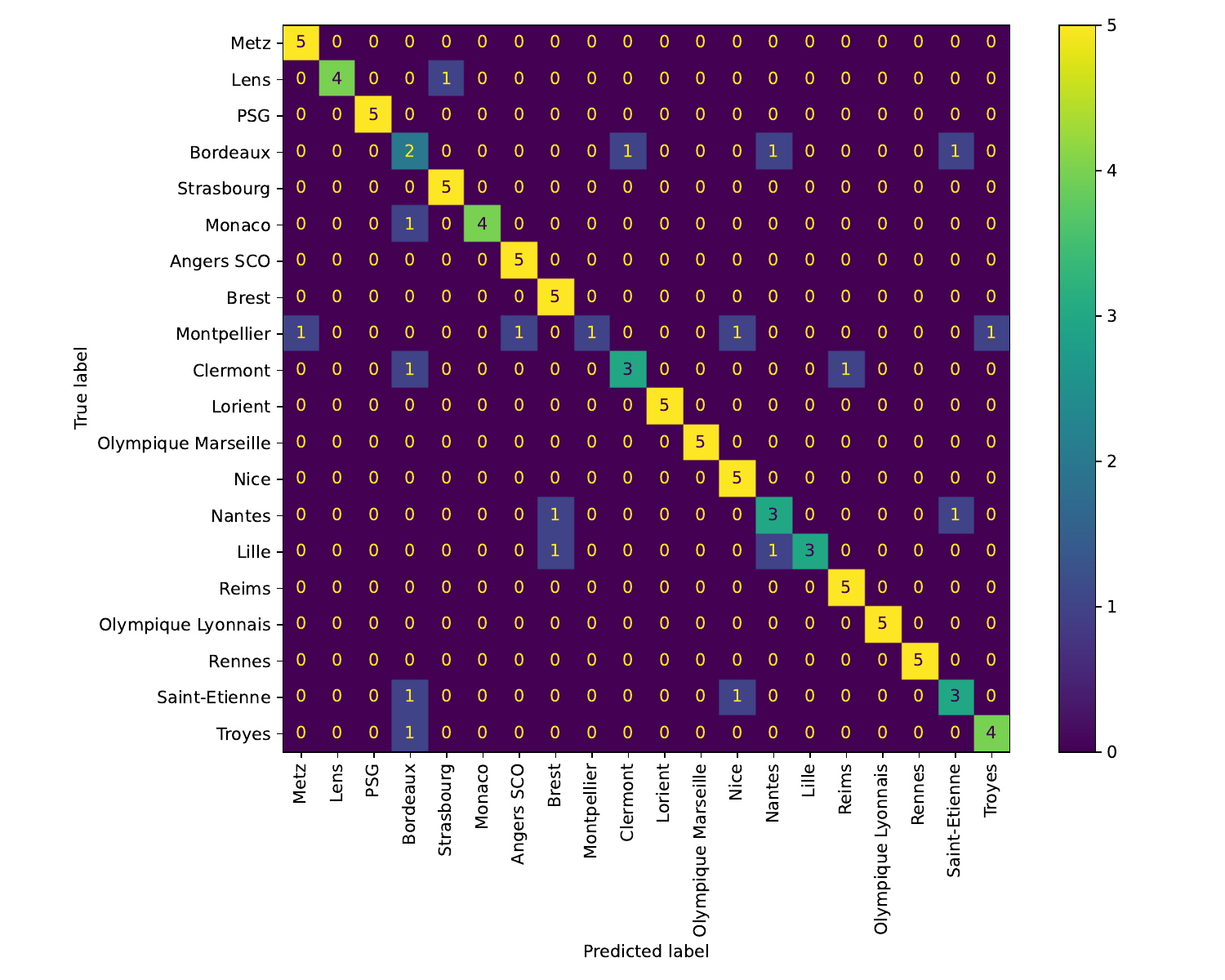}}
    \caption{Confusion matrix of the Gaussian Mixture Model with 50 components. The matrix illustrates correct and incorrect predictions for each team across the $5$ folds.}
    \label{fig:confusion_matrix}
\end{figure}

\begin{figure}[t]
    \centering
    \includegraphics[width=0.8\textwidth]{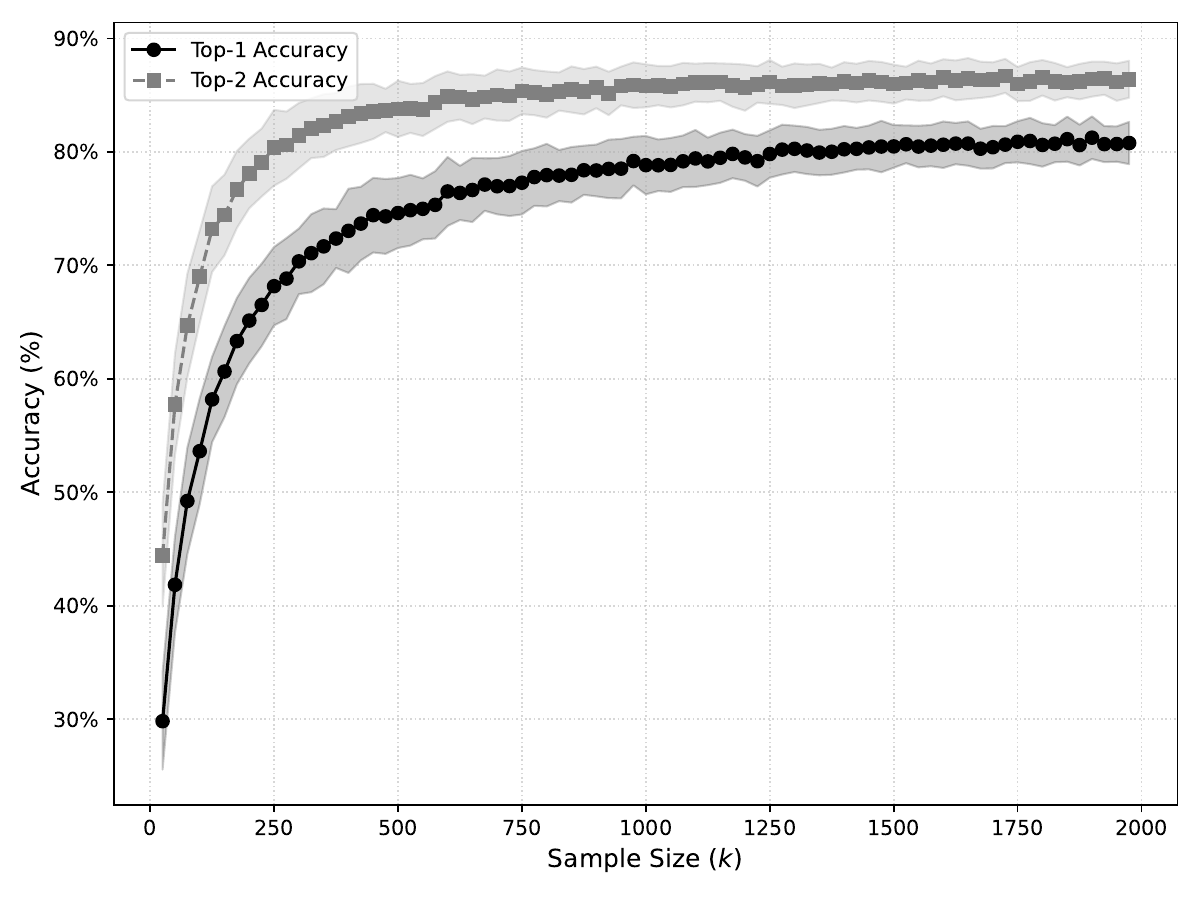}
    \caption{Accuracy of team identity prediction as a function of test sample size. For each sample size, $100$ randomly selected subsets of the test fold are used to compute the accuracy and its standard deviation.}
    \label{fig:accuracy_sample_size}
\end{figure}

\noindent
In this section, we demonstrate that our representation of collections of frames as distributions in the embedding space effectively captures the stylistic identity of teams. Specifically, we show that it is possible to predict a team's identity from out-of-sample frames with high accuracy, which provides strong evidence that our embedding preserves spatial information while differentiating between unique playing styles. The task we try to solve is to predict the identity of a team using as input a collection of frames from the games of this team. To achieve this goal, we learn the distribution of the frames of each team. Then, given a new collection of frames, we compute the likelihood of those frames with respect to the learned distribution of each team and select the maximum a posteriori as the predicted identity of the team. 
\\
\noindent
\\
To learn the distribution of frames of a team, we propose the use of a spherical Gaussian mixture. The embedding we use maps frames to the large dimensional space $\mathbb{R}^{11\times 12}$, which makes modeling with a regular Gaussian mixture computationally expensive. In fact, the Expectation-Maximization algorithm requires the storage and inversion of large covariance matrices for each component. For this reason, we choose to fit a spherical Gaussian mixture model. This has the added advantage that the fitted Gaussian components are isotropic around their centers, depending only on the $\mathbb{L}^2$ distance. This aligns with our initial intuition that the $\mathbb{L}^2$ distance between embeddings corresponds to an optimal transport cost between frames. In particular, the spherical GMM modeling assumes that the likelihood depends only on the sliced-Wasserstein distance to the component means. 
\\
\noindent
\\
To evaluate this methodology, we employ a 5-fold cross-validation procedure. Let \(\mathcal{C}_t\) denote the collection of frames for team $t$, embedded in the space $\mathbb{R}^{n \times L}$. Each collection $\mathcal{C}_t$ is divided into five consecutive folds, denoted by $\mathcal{C}_t^{(l)}$ for $l = 1, \dots, 5$. For each fold $l = 1, \dots, 5$, we perform the following steps:

\begin{enumerate}
    \item \textbf{Training Phase}:  
    Exclude the $l^{\text{th}}$ fold from each team's dataset, resulting in the training set:
    \[
    \mathcal{C}_t^{(-l)} = \mathcal{C}_t \setminus \mathcal{C}_t^{(l)}.
    \]
    We then fit a spherical Gaussian Mixture Model with 50 components to the training data $\mathcal{C}_t^{(-l)}$ for each team using the Expectation-Maximization (EM) algorithm.
    
    \item \label{enumerate:testphase} \textbf{Testing Phase}:  
    For the held-out fold $\mathcal{C}_t^{(l)}$ of each team $t$, we predict the team identity by calculating the likelihood of the collection of frames under the GMM of every team $t'$. 
    \item \textbf{Classification Evaluation}:  
    \begin{itemize}
        \item \textit{Top-1 Classification}: We record a Top-1 classification if the true team identity corresponds to the highest likelihood GMM.
        \item \textit{Top-2 Classification}: We record a Top-2 classification if the true team identity is among the two highest likelihood GMMs.
    \end{itemize}
\end{enumerate}
\noindent
After completing all five folds, we aggregate the classification results. In this step, we evaluate the ability of the methodology in predicting the team identity using the entire test fold collection of frames. This yields an average Top-1 accuracy of 82\% and a Top-2 accuracy of 88\%. Figure~\ref{fig:confusion_matrix} shows the confusion matrix for these predictions. Teams with more distinctive playing styles tend to have higher classification accuracy, whereas teams with more balanced or less distinctive styles exhibit slightly lower performance. This is consistent with the results in Table~\ref{tab:sum_distances_team}. In particular, the teams that the methodology struggles to distinguish are Montpellier and Bordeaux and both show small distances to the other teams using our similarity metric. This pattern suggests that the embedding encodes meaningful tactical behaviors and spatial patterns associated with each team's approach.
\\
\noindent
\\
This methodology aims at predicting the identity of a team given a collection of their frames. Therefore, we are interested in evaluating the impact of the size of such sample on the accuracy of the prediction. To do so, for each sample size $k$,  we randomly subsample the held-out fold $\mathcal{C}_t^{(-l)}$ in Step~\ref{enumerate:testphase} keeping a subset of $k$ frames. We then predict the team identity and calculate the mean accuracy across teams and folds. This experiment is repeated $100$ times for each sample size $k$, yielding a measure of uncertainty through the standard deviation of the estimated mean accuracy. This allows us to to compute mean accuracy and standard deviation, highlighting the robustness of the classifier at different sampling levels. Figure~\ref{fig:accuracy_sample_size} shows how accuracy varies as a function of the number of frames in each subset. Larger subsets naturally improve accuracy, as they offer a more representative idea of a team's positioning style. Furthermore, we observe that as much as $300$ frames are sufficient to achieve $70.35\%$ Top-1 accuracy and $81.43\%$ in predicting the correct team among the two most likely. Overall, the results confirm that our embedding captures team-specific spatial configurations, enabling accurate team identification even with a relatively small number of frames.

\clearpage
\section{Conclusion}
 This study presents a novel framework for analyzing and comparing the playing styles of football teams, focusing on the positional configurations of players during matches. By representing collections of frames as distributions in an embedding space, our methodology provides an efficient and scalable approach to capturing team-specific spatial behavior, revealing meaningful insights into team tactics and style. We apply our methodology to games from the 2021-2022 Ligue 1 season to determine a similarity matrix between the collections of frames of each team. We retrieve values that are consistent with the possession values in each collection of games. These values enable us to compare the styles of play displayed by the teams in each game. Furthermore, the quantization step can be interpreted as a clustering algorithm and used to determine groups of game situations observed in a given team's games. 
 \\
\noindent
\\
 One of the key aspects of this work is evaluating the ability of our embedding to capture the stylistic identity of teams, which we demonstrate through the successful prediction of team identity using a Bayesian classifier. The classifier achieved a Top-1 accuracy of 82\% and a Top-2 accuracy of 88\%, confirming that our embedding effectively differentiates between teams based on their spatial patterns. Further, we explored the impact of sample size on the classifier's performance, finding that even a relatively small subset of $300$ frames was sufficient to achieve high accuracy. This demonstrates the robustness and efficiency of our embedding in capturing essential team features, even with limited data. The embedding was also successfully evaluated in a classification task to predict team possession given player locations. The optimal transport based embedding provides a preprocessing tool to feed spatial information from a frame to a machine learning model. The embedding is permutation invariant and preserves the distance between frames. 
 \\
\noindent
\\
 In addition to its application in football, we extend our framework to analyze the playing styles of NBA teams, showcasing its versatility across sports. The ability to generalize the methodology and apply it to different datasets and to solve different tasks opens up new possibilities for cross-sport analysis and tactical optimization.
 \\
 \noindent
 \\
 In this work, we have introduced tools to extract and leverage spatial information from both individual frames and collections of frames. Their applications range from quantifying similarity to solving prediction tasks. Among the various ways these tools can benefit the football industry, scouting stands out as a key application. In particular, measuring the similarity between teams based on their spatial distributions can provide critical insights into the potential success of a prospective transfer. By contextualizing  a player's attributes within a team's spatial strategy, scouts can better predict the success of a transfer and make informed recruitment decisions. Another class of applications includes offering managers tools based on the embedding. For example, It allows the easy identification of situations in a collection frames that are spatially close to a given scenario, specified by the coach.
\\
\noindent
\\
\textbf{Acknowledgment}: The authors thank Stats Perform, Matthieu Lille-Palette and Andy Cooper for providing tracking data. They are also grateful to Mathieu Lacome and Sébastien Coustou for insightful discussions. The
authors gratefully acknowledge financial support from the chair “Machine Learning \& Systematic
Methods in Finance” from Ecole Polytechnique.

\clearpage
\bibliographystyle{apalike}
\bibliography{biblio}

\newpage
\newpage
\begin{center}
{\Large \bf {\centering Appendix}}
\end{center}
\appendix
\label{appendix}
\section{Distance between teams in the NBA}
\begin{figure}[H]
    \centering
    \begin{minipage}{0.9\textwidth}
        \centering
        \scalebox{0.40}{\includegraphics{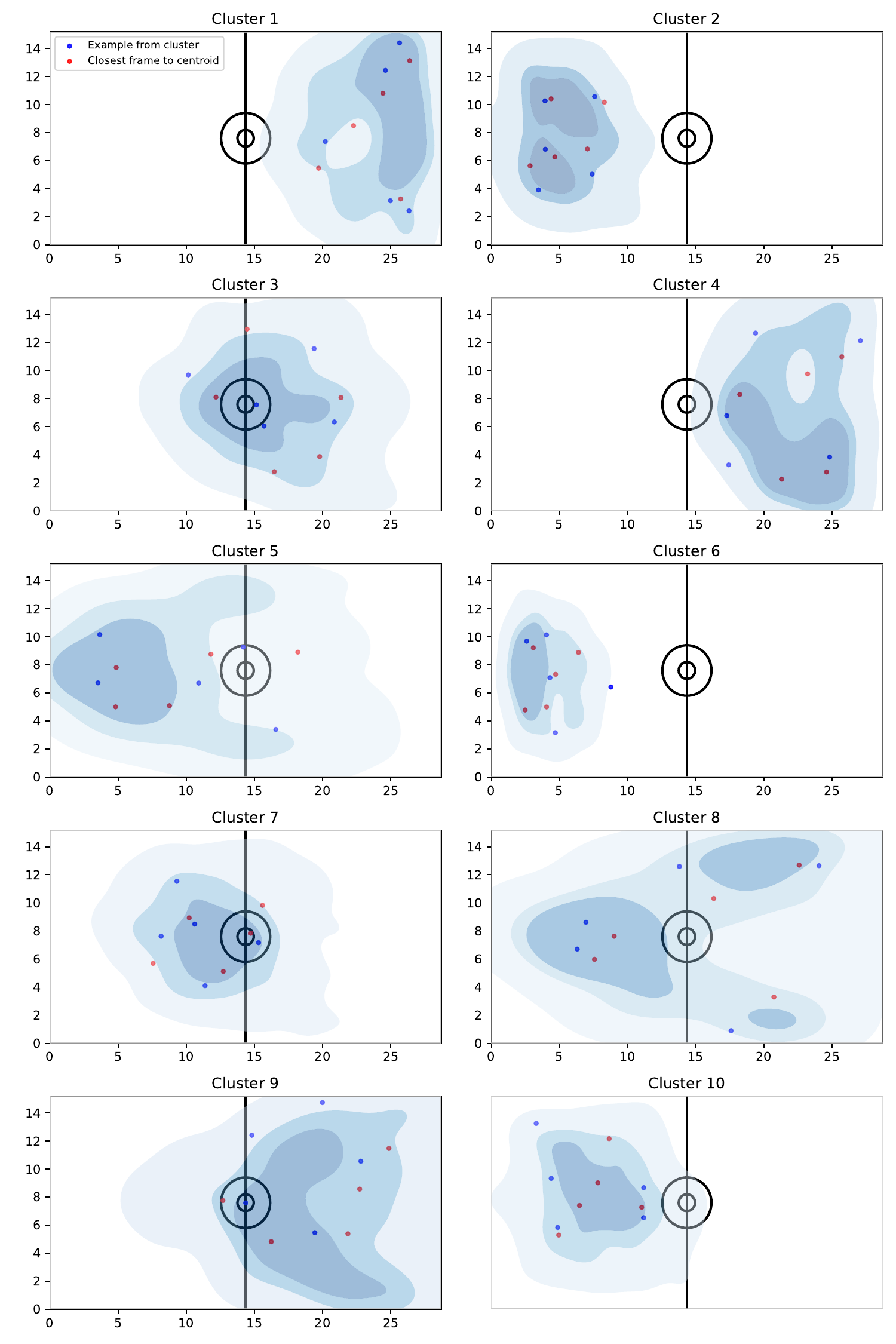}}\vspace*{-.5em}
        \captionof{figure}{Example of $10$ clusters observed in the frames from the games of the Golden State Warriors. For each cluster, the closest frame to the barycenter and a random example from the cluster are shown.  The shaded blue color represents the total density of player locations taken from a random sample of 100 frames from each cluster.}\label{fig:graphnba}      
    \end{minipage}
\end{figure}

Similarly to Section \ref{section:results}, we apply the optimal transport framework to tracking data from the first half of the 2015-2016 NBA season. For each team, the tracking data was subsampled by retaining one out of every 25 frames. We set $L=6$ to obtain an embedding in $\mathbb{R}^{5 \times 6}$ by projecting along the grid of directions defined for $l=1,\dots,L$ by
$$
\theta_l = \left( \cos\left(\frac{\pi (l-1)}{2L}\right), \sin\left(\frac{\pi (l-1)}{2L}\right) \right).
$$
\\
\noindent
\\
Figure \ref{fig:graphnba} illustrates an example of the ten clusters identified from the collection of frames across all Golden State Warriors games, and Table~\ref{tab:occurencegsw} provides the percentage of frames in each cluster. We observe that the clustering algorithm effectively distinguishes different phases of play in a consistent manner. Specifically, we identify three distinct defensive settings in clusters 6, 2, and 10, where the defensive block transitions from very deep positions to more advanced ones. Additionally, clusters 3, 5, 8, and 9 represent transitions between defense and offense, characterized by players being spread across the court. The dynamics observed in these clusters differ significantly from those in football. In particular, clusters where players are positioned in the center of the court account for a total of 18.62\% of the frames, unlike in football, where central clusters are more prevalent. This difference is expected because basketball rules prohibit the ball from moving backward from the offensive half of the court. Consequently, basketball is essentially a transition-oriented game where teams alternate between defense and offense.

\begin{table}[H]
    \centering
    \resizebox{0.3\columnwidth}{!}{
    \begin{tabular}{ll}
        \toprule
        \toprule
        \textbf{Cluster} & \textbf{Percentage of frames} \\ 
        \midrule
        1  & 13.04\% \\
        2  & 22.49\% \\
        3  & 4.73\%  \\
         \midrule
        4  & 15.66\% \\
        5  & 3.40\%  \\
        6  & 17.31\% \\
         \midrule
        7  & 5.22\%  \\
        8  & 3.27\%  \\
        9  & 7.22\%  \\
        10 & 7.66\%  \\
        \bottomrule
        \bottomrule
    \end{tabular}}
    \vspace{0.5em}
    \caption{Occurrence of clusters displayed in Figure \ref{fig:graphnba}.}
    \label{tab:occurencegsw}
\end{table}

\begin{figure}[H]
    \centering
    \includegraphics[width=0.8\textwidth]{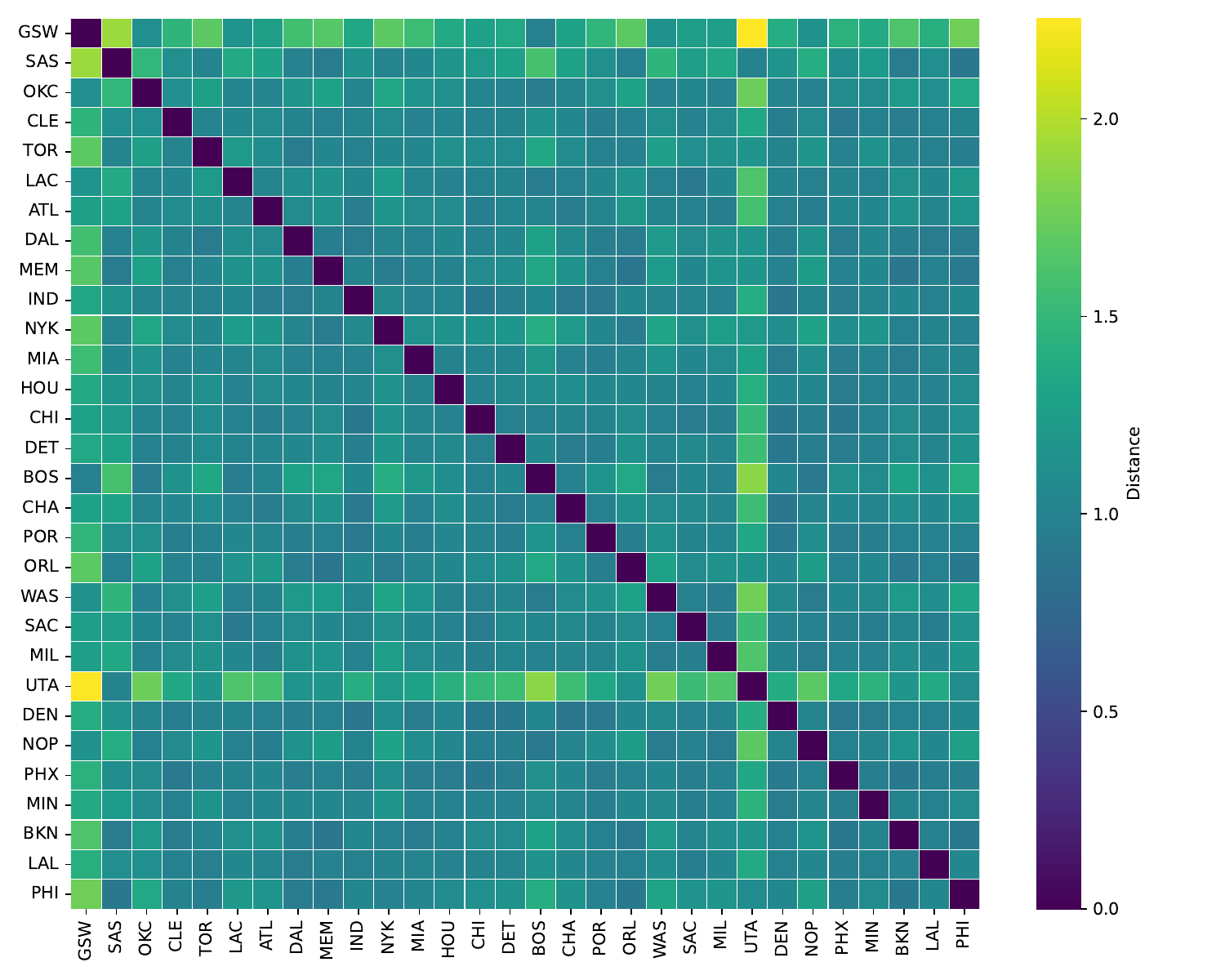}   
    \caption{Distance matrix between the collections of frames in the games of each team in the NBA.}\label{fig:similanba}
\end{figure}

\begin{figure}[H]
    \centering
    \includegraphics[width=0.8\textwidth]{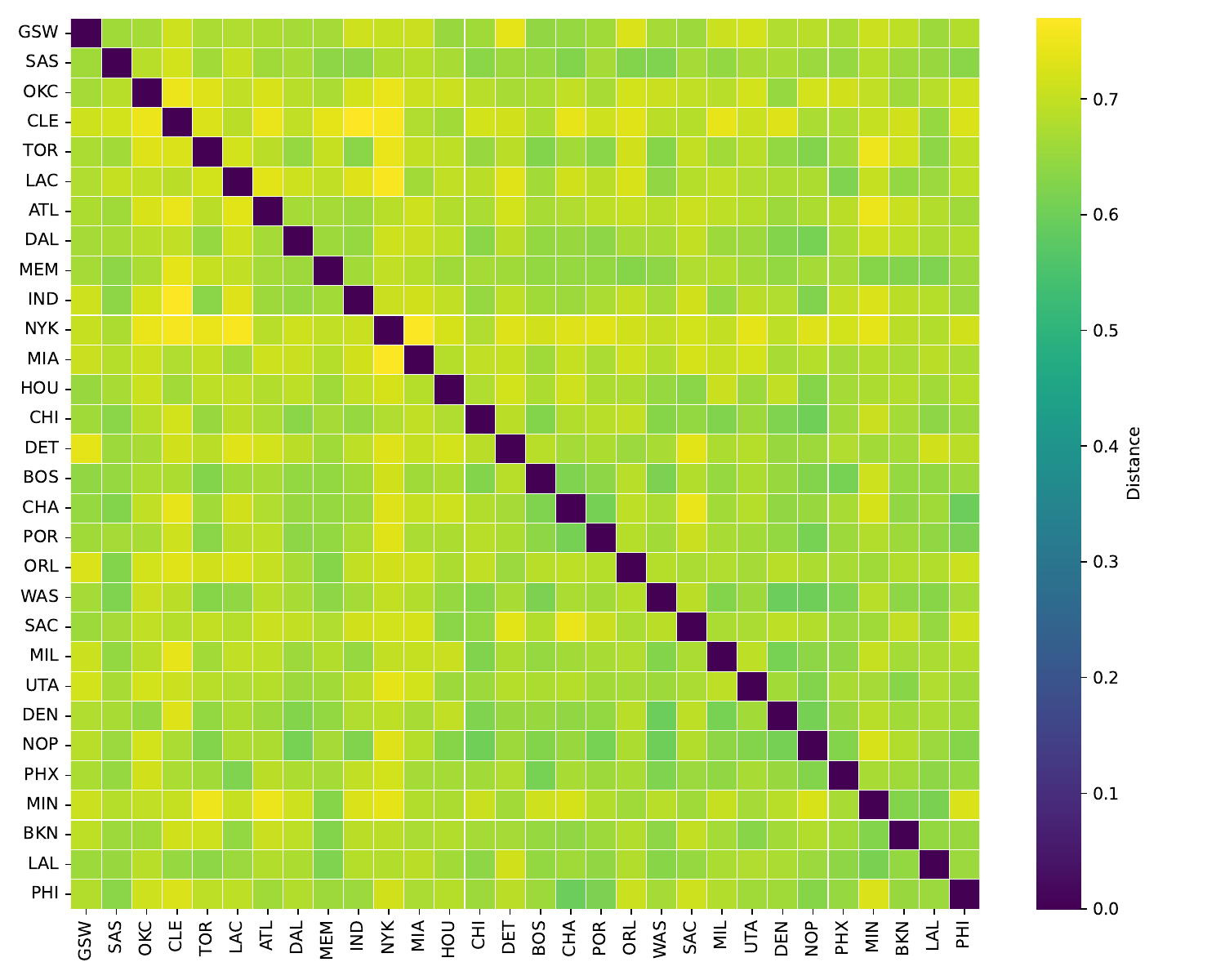}   
    \caption{Distance matrix between the collections of frames in the games of each team in the NBA after centering the data.}\label{fig:similanbacentered}
\end{figure}

\noindent
Finally, Figure \ref{fig:similanba} illustrates the similarity in playing styles between NBA teams. This similarity is computed by considering the collection of frames from all available games in the first half of the NBA season, subsampled at a rate of one frame every 25 frames. We observe that the Golden State Warriors exhibit the largest deviation from the other franchises. This is not surprising, as they are credited with introducing a new style of play heavily reliant on 3-pointers. In fact, this season marked the record for the most points scored in the regular season, which naturally leads to a different spatial distribution of play. Interestingly, the Utah Jazz display the second largest deviation from the rest of the teams and its distance to the Golden State Warriors is the greatest. Figure \ref{fig:similanbacentered} presents the similarity metric after centering the frames prior to embedding. We observe in this case that the distance metrics become more uniform, and the significant deviations of the Utah Jazz and Golden State Warriors diminish. This suggests that these effects are primarily due to the average locations both teams occupy during their games rather than relative placement of players.


\subsection{Predicting Team Identity}
\begin{figure}[H]
    \centering
    \includegraphics[width=0.8\textwidth]{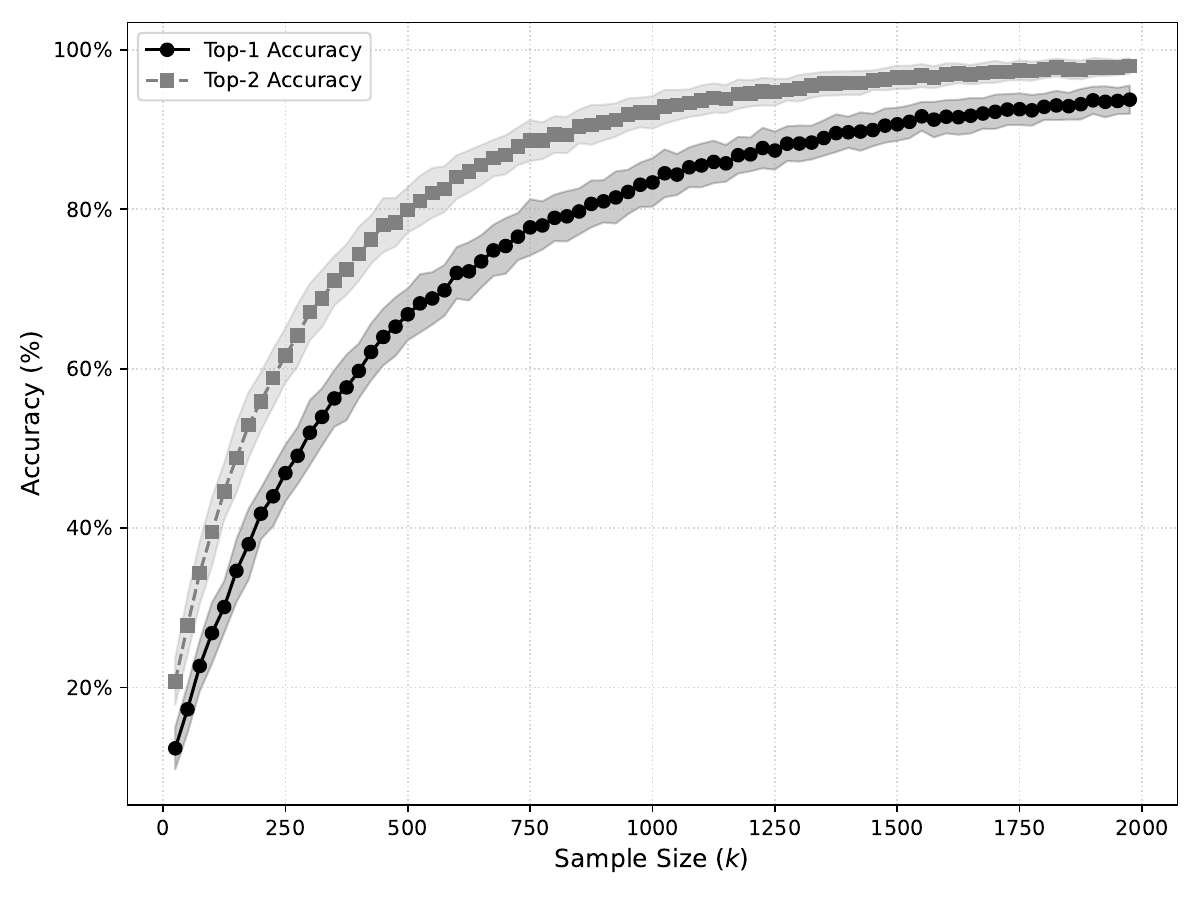}
    \caption{Accuracy of NBA team identity prediction as a function of test sample size. For each sample size, $100$ randomly selected subsets of the test fold are used to compute the accuracy and its standard deviation.}
    \label{fig:nba_accuracy_sample_size}
\end{figure}

\noindent
Similarly to Section \ref{sec:teamidentity}, we demonstrate that our representation of collections of frames as distributions in the embedding space effectively captures the stylistic identity of teams in the NBA. Specifically, we show that it is possible to predict a team's identity from an out-of-sample collection of frames with high accuracy, which provides strong evidence that our embedding preserves spatial information while differentiating between unique playing styles.
\\
\noindent
\\
Using all frames in each test fold yields an average Top-1 accuracy of 99.33\% and a Top-2 accuracy of 100\%. This performance suggests that the distribution of frames in the embedding perfectly captures the style of play of a team with a large enough test sample of frames. Figure~\ref{fig:nba_accuracy_sample_size} shows how accuracy varies as a function of the number of frames in each subset. Similarly to the football data, larger subsets naturally improve accuracy, as they offer a more representative idea of a team's positioning style. Compared to football, we need more frames to achieve an accuracy of $70\%$, but the ceiling of precision is higher. In fact, we achieve $93.8\%$ Top-1 accuracy with $2000$ frames. Overall, the results further confirm that our embedding captures team-specific spatial configurations, enabling accurate and fast team identification.

\newpage
\section{Proof of Proposition \ref{lemma:distance}}

First, we prove that $\text{Proj}_{\theta}$ is injective. We proceed by induction on $n$
\\
\noindent
\\
 $(H_n)$: For any choice of grid $\theta_1,\theta_2,\dots,\theta_L$ such that $L\geq n+1$ and $\theta_l,\theta_k$ are noncollinear for all $l,k\leq L$, $\text{Proj}_{\theta}$ is injective. 
\\
\noindent
\\
If $n=1$, take $\mu=\delta_x$ and $\nu=\delta_y$ and $\theta_1,\theta_2,\dots,\theta_L$ such that $L\geq 2$ 
 and $\text{Proj}_{\theta}(\mu)=\text{Proj}_{\theta}(\nu)$. Then we have $\delta_{\langle\theta_1,x\rangle}=\delta_{\langle\theta_1,y\rangle}$ and $\delta_{\langle\theta_2,x\rangle}=\delta_{\langle\theta_2,y\rangle}$. Equivalently, $\langle\theta_1,x\rangle=\langle\theta_1,y\rangle$ and $\langle\theta_2,x\rangle=\langle\theta_2,y\rangle$ which implies $x=y$ since $\theta_1$ and $\theta_2$ are non collinear. Thus, $\mu=\nu$.
\\
\noindent
\\
Let $n\geq 2$ and assume $(H_{n-1})$ is true. Take $\mu$ and $\nu$ in $\mathcal{P}_n^{\text{u}}(\mathbb{R}^2)$ such that $\text{Proj}_{\theta}(\mu)=\text{Proj}_{\theta}(\nu)$.
We can write $\mu=\sum_{i=1}^{n}\frac{1}{n}\delta_{x_{i}}$ and $\nu=\sum_{i=1}^{n}\frac{1}{n}\delta_{y_{i}}$ and we have for all $l$ in $\{1,\dots,L\}$:
\begin{equation*}
\sum_{i=1}^{n}\frac{1}{n}\delta_{\langle\theta_l,x_{i}\rangle}=\sum_{i=1}^{n}\frac{1}{n}\delta_{\langle\theta_l,y_{i}\rangle}.
\end{equation*}
Therefore, for all $l$ in $\{1,\dots,L\}$, there exists $i_l$ in $ \{1,\dots,n\}$ such that $$\langle\theta_l,x_{n}\rangle=\langle\theta_l,y_{i_l}\rangle.$$
Since $L\geq n+1$ and using the pigeon-hole principle, there must exist $l,k$ such that $i_l=i_k$. Thus 
\begin{align*}
 &\langle\theta_l,x_{n}\rangle=\langle\theta_l,y_{i_l}\rangle,\\
&\langle\theta_k,x_{n}\rangle=\langle\theta_k,y_{i_l}\rangle .
\end{align*}
we can deduce that $x_{n}=y_{i_l}$. And we have \begin{equation*}
\sum_{i=1}^{n-1}\frac{1}{n-1}\delta_{\langle\theta_l,x_{h,i}\rangle}=\sum_{i\neq i_l}\frac{1}{n-1}\delta_{\langle\theta_l,y_{h,i}\rangle}.
\end{equation*}
We conclude using $(H_{n-1})$ on $\mu'=\sum_{i=1}^{n-1}\frac{1}{n-1}\delta_{x_{h,i}}$ and $\nu'=\sum_{i\neq i_l}\frac{1}{n-1}\delta_{y_{h,i}}$.
\\
\noindent
\\
Finally, for $\mu$ and $\nu$ in $\mathcal{P}_n^{\text{u}}(\mathbb{R}^2)$ we have \begin{equation*}
    \widehat{SW}_p(\mu,\nu,(\theta_l)_{l\leq L})= \frac{1}{(nL)^{1/p}}\|\text{Proj}_{\theta}(\mu)-\text{Proj}_{\theta}(\nu) \|_p.
\end{equation*}
This justifies that $\widehat{SW}_p(.,.,(\theta_l)_{l\leq L})$ is a distance over $\mathcal{P}_n^{\text{u}}(\mathbb{R}^2)$.

\newpage

\section{Justification of the normalisation term}
The following result allows to determine a suitable normalisation factor for the sliced-Wasserstein distance.
\begin{proposition}
\label{prop:equivalence}
For $\mu, \nu \in \mathcal{P}^{\text{u}}_{n}(\mathbb{R}^2)$, we have
\begin{equation}
\label{eq:bound1}
\widehat{SW}_2(\mu, \nu, (\theta_{l})_{l \leq L}) \leq \sqrt{\frac{1 + \frac{1}{L \sin\left(\frac{\pi}{2L}\right)}}{2}} W_2(\mu, \nu).
\end{equation}
\noindent
Furthermore, in the case where $\nu = \delta_y$ for $y \in \mathbb{R}^2$:
\begin{equation}
\label{eq:bound2}
\widehat{SW}_2(\mu, \nu, (\theta_{l})_{l \leq L}) \geq \sqrt{\frac{1 - \frac{1}{L \sin\left(\frac{\pi}{2L}\right)}}{2}} W_2(\mu, \nu).
\end{equation}
\end{proposition}
\noindent
To prove Equation \eqref{eq:bound1}, consider $\mu=\frac{1}{n}\sum_{i=1}^n \delta_{x_i}$ and $\nu=\frac{1}{n}\sum_{i=1}^n \delta_{y_i}$ and let $\sigma$ be a permutation such that 
\begin{equation*}
    W_2^2(\mu,\nu)=  \frac{1}{n}\sum_{i=1}^n \|x_i-y_{\sigma(i)}\|^2.
\end{equation*}
Then, for all $l=1,\dots,L$, we have
\begin{equation*}
    W_2^2(\theta_l\#\mu,\theta_l\#\nu)\leq   \frac{1}{n}\sum_{i=1}^n \left |\langle\theta_l,x_i\rangle- \langle\theta_l,y_{\sigma(i)}\rangle \right |^2.
\end{equation*}
Thus, 
\begin{align*}
     \widehat{SW}_2(\mu,\nu,(\theta_{l})_{l\leq L})&\leq   
     \left ( \frac{1}{nL} \sum_{l=1}^L \sum_{i=1}^n \left | \langle\theta_l,x_{i}-y_{\sigma(i)}\rangle  \right |^2 \right )^{1/2}, \\
     &=\left ( \frac{1}{nL}\sum_{i=1}^n \sum_{l=1}^L   (x_i-y_{\sigma(i)})^T\theta_l\theta_l^T(x_{i}-y_{\sigma(i)})  \right )^{1/2},\\
    &=\left ( \frac{1}{nL}\sum_{i=1}^n    (x_i-y_{\sigma(i)})^T\Theta (x_{i}-y_{\sigma(i)})  \right )^{1/2},
\end{align*}
where $\Theta = \sum\limits_{l=1}^L \theta_l\theta_l^T$ is a positive semi-definite matrix. Let $\rho(\Theta)$ be its spectral radius, we have
\begin{align*}
     \widehat{SW}_2(\mu,\nu,(\theta_{l})_{l\leq L})& \leq \sqrt{\rho(\Theta)} \left ( \frac{1}{nL}\sum_{i=1}^n   \|x_{i}-y_{\sigma(i)}\|^2 \right )^{1/2},\\
     &=\sqrt{\frac{\rho(\Theta)}{L}} W_2(\mu,\nu).
\end{align*}
\noindent
For the choice of grid in Equation \eqref{eq:grid}, we have $\theta_l\theta_l^T=\begin{pmatrix}
\cos^2(\frac{\pi (l-1) }{2L}) & \cos(\frac{\pi (l-1) }{2L})\sin(\frac{\pi (l-1) }{2L}) \\
\cos(\frac{\pi (l-1) }{2L})\sin(\frac{\pi (l-1) }{2L}) & \sin^2(\frac{\pi (l-1) }{2L}) \\
\end{pmatrix}$ and hence
\begin{align*}
    \Theta&= \begin{pmatrix}
\frac{L-1}{2} & \frac{\cos(\frac{\pi }{2L})}{2\sin(\frac{\pi }{2L})} \\
\frac{\cos(\frac{\pi }{2L})}{2\sin(\frac{\pi }{2L})}& \frac{L+1}{2} \\
\end{pmatrix}.
\end{align*}
This yields $\rho(\Theta)=\frac{L}{2}(1+\frac{1}{L\sin(\frac{\pi}{2L}}))$ and we deduce that 
\begin{align*}
     \widehat{SW}_2(\mu,\nu,(\theta_{l})_{l\leq L})&\leq \sqrt{\frac{1+\frac{1}{L\sin(\frac{\pi}{2L})}}{2}} W_2(\mu,\nu).
\end{align*}
 For the second bound, we consider the case where  $\nu=\delta_y$ is concentrated in one location. Similar calculations yield
\begin{align*}
     \widehat{SW}_2(\mu,\nu,(\theta_{l})_{l\leq L})&=   
     \left ( \frac{1}{nL} \sum_{l=1}^L \sum_{i=1}^n \left | \langle\theta_l,x_{i}-y\rangle  \right |^2 \right )^{1/2}, \\
    &=\left ( \frac{1}{nL}\sum_{i=1}^n    (x_i-y)^T\Theta (x_{i}-y)  \right )^{1/2},\\
    &\geq \sqrt{\frac{\gamma(\Theta)}{L}} \left ( \frac{1}{n}\sum_{i=1}^n   \|x_{i}-y\|^2 \right )^{1/2},\\
\end{align*}
where $\gamma(\Theta)$ is the smallest eigenvalue of $\Theta$ and is given, for the grid in Equation \eqref{eq:grid}, by $\gamma(\Theta)= \frac{L}{2}(1-\frac{1}{L\sin(\frac{\pi }{2L})} )$. Therefore, we obtain
\begin{align*}
     \widehat{SW}_2(\mu,\nu,(\theta_{l})_{l\leq L})&\geq   
     \sqrt{\frac{1-\frac{1}{L\sin(\frac{\pi}{2L})}}{2}} W_2(\mu,\nu).
\end{align*}

\newpage

\begin{table}[H]
    \centering
    \resizebox{0.4\columnwidth}{!}{\begin{tabular}{ll}
        \toprule
        \toprule
        \textbf{Team} & \textbf{Possession Value (\%)} \\
        \midrule
        Olympique Marseille &   63.85\% \\
PSG &                   60.63\% \\
Rennes &                60.15\% \\
\midrule
Olympique Lyonnais &    58.36\% \\
Nice &                  55.15\% \\
Lille &                 54.77\% \\
\midrule
Lens &                  51.05\% \\
Clermont &              49.30\% \\
Saint-Étienne &        48.71\% \\
\midrule
Montpellier &           48.32\% \\
Bordeaux &              48.29\% \\
Strasbourg &            47.72\% \\
\midrule
Brest &                 46.87\% \\
Monaco &                46.62\% \\
Angers SCO &            45.71\% \\
\midrule
Metz &                  45.19\% \\
Nantes &                45.02\% \\
Troyes &                44.22\% \\
\midrule
Lorient &               43.43\% \\
Reims &                 41.63\% \\
        \bottomrule
        \bottomrule
    \end{tabular}}
    \vspace{0.5em}
    \caption{Possession values for each team in the analysis, displayed as percentages. Teams are sorted in descending order.}
    \label{tab:team_metrics_percentages}
\end{table}

\begin{table}[ht]
    \centering
    \begin{subtable}[t]{0.28\textwidth}
        \centering
        \tiny
        \resizebox{\linewidth}{!}{%
            \begin{tabular}{rlll}
                \toprule
                \toprule
                \textbf{Game} & \textbf{Home Team} & \textbf{Away Team} & \textbf{Date} \\
                \midrule
                1 & Metz & Lens & 2022-03-13 \\
                2 & PSG & Bordeaux & 2022-03-13 \\
                3 & Strasbourg & Monaco & 2022-03-13 \\
                \midrule
                4 & Angers SCO & Brest & 2022-03-20 \\
                5 & Bordeaux & Montpellier & 2022-03-20 \\
                6 & Lens & Clermont & 2022-03-19 \\
                \midrule
                7 & Lorient & Strasbourg & 2022-03-20 \\
                8 & Olympique Marseille & Nice & 2022-03-20 \\
                9 & Monaco & PSG & 2022-03-20 \\
                \midrule
                10 & Nantes & Lille & 2022-03-19 \\
                11 & Reims & Olympique Lyonnais & 2022-03-20 \\
                12 & Rennes & Metz & 2022-03-20 \\
                \midrule
                13 & Saint-Etienne & Troyes & 2022-03-18 \\
                14 & Clermont & Nantes & 2022-04-03 \\
                15 & Lille & Bordeaux & 2022-04-02 \\
                \midrule
                16 & Olympique Lyonnais & Angers SCO & 2022-04-03 \\
                17 & Metz & Monaco & 2022-04-03 \\
                18 & Montpellier & Brest & 2022-04-03 \\
                \midrule
                19 & Nice & Rennes & 2022-04-02 \\
                20 & PSG & Lorient & 2022-04-03 \\
                21 & Saint-Etienne & Olympique Marseille & 2022-04-03 \\
                \midrule
                22 & Strasbourg & Lens & 2022-04-03 \\
                23 & Troyes & Reims & 2022-04-03 \\
                24 & Angers SCO & Lille & 2022-04-10 \\
                \midrule
                25 & Brest & Nantes & 2022-04-10 \\
                26 & Bordeaux & Metz & 2022-04-10 \\
                27 & Clermont & PSG & 2022-04-09 \\
                \midrule
                28 & Lens & Nice & 2022-04-10 \\
                29 & Lorient & Saint-Etienne & 2022-04-08 \\
                30 & Monaco & Troyes & 2022-04-10 \\
                \midrule
                31 & Reims & Rennes & 2022-04-09 \\
                32 & Strasbourg & Olympique Lyonnais & 2022-04-10 \\
                33 & Lille & Lens & 2022-04-16 \\
                \bottomrule
                \bottomrule
            \end{tabular}
        }
    \end{subtable}
    \hfill
    \begin{subtable}[t]{0.28\textwidth}
        \centering
        \tiny
        \resizebox{\linewidth}{!}{%
            \begin{tabular}{rlll}
                \toprule
                \toprule
                \textbf{Game} & \textbf{Home Team} & \textbf{Away Team} & \textbf{Date} \\
                \midrule
                34 & Olympique Lyonnais & Bordeaux & 2022-04-17 \\
                35 & Metz & Clermont & 2022-04-17 \\
                36 & Nantes & Angers SCO & 2022-04-17 \\
                \midrule
                37 & Nice & Lorient & 2022-04-17 \\
                38 & PSG & Olympique Marseille & 2022-04-17 \\
                39 & Rennes & Monaco & 2022-04-15 \\
                \midrule
                40 & Saint-Etienne & Brest & 2022-04-16 \\
                41 & Troyes & Strasbourg & 2022-04-17 \\
                42 & Angers SCO & PSG & 2022-04-20 \\
                \midrule
                43 & Bordeaux & Saint-Etienne & 2022-04-20 \\
                44 & Lens & Montpellier & 2022-04-20 \\
                45 & Lorient & Metz & 2022-04-20 \\
                \midrule
                46 & Olympique Marseille & Nantes & 2022-04-20 \\
                47 & Monaco & Nice & 2022-04-20 \\
                48 & Reims & Lille & 2022-04-20 \\
                \midrule
                49 & Strasbourg & Rennes & 2022-04-20 \\
                50 & Troyes & Clermont & 2022-04-20 \\
                51 & Brest & Olympique Lyonnais & 2022-04-20 \\
                \midrule
                52 & Lille & Strasbourg & 2022-04-24 \\
                53 & Olympique Lyonnais & Montpellier & 2022-04-23 \\
                54 & Metz & Brest & 2022-04-24 \\
                \midrule
                55 & Nantes & Bordeaux & 2022-04-24 \\
                56 & Nice & Troyes & 2022-04-24 \\
                57 & PSG & Lens & 2022-04-23 \\
                \midrule
                58 & Reims & Olympique Marseille & 2022-04-24 \\
                59 & Rennes & Lorient & 2022-04-24 \\
                60 & Saint-Etienne & Monaco & 2022-04-23 \\
                \midrule
                61 & Olympique Marseille & Olympique Lyonnais & 2022-05-01 \\
                62 & Strasbourg & PSG & 2022-04-29 \\
                63 & Angers SCO & Bordeaux & 2022-05-08 \\
                \midrule
                64 & Clermont & Montpellier & 2022-05-08 \\
                65 & Lille & Monaco & 2022-05-06 \\
                66 & Metz & Olympique Lyonnais & 2022-05-08 \\
                \bottomrule
                \bottomrule
            \end{tabular}
        }
    \end{subtable}
    \hfill
    \begin{subtable}[t]{0.28\textwidth}
        \centering
        \tiny
        \resizebox{\linewidth}{!}{%
            \begin{tabular}{rlll}
                \toprule
                \toprule
                \textbf{Game} & \textbf{Home Team} & \textbf{Away Team} & \textbf{Date} \\
                \midrule
                67 & Nantes & Rennes & 2022-05-11 \\
                68 & Nice & Saint-Etienne & 2022-05-11 \\
                69 & PSG & Troyes & 2022-05-08 \\
                \midrule
                70 & Reims & Lens & 2022-05-08 \\
                71 & Metz & Angers SCO & 2022-05-14 \\
                72 & Monaco & Brest & 2022-05-14 \\
                \midrule
                73 & Montpellier & PSG & 2022-05-14 \\
                74 & Nice & Lille & 2022-05-14 \\
                75 & Rennes & Olympique Marseille & 2022-05-14 \\
                \midrule
                76 & Bordeaux & Nice & 2022-05-01 \\
                77 & Brest & Clermont & 2022-05-01 \\
                78 & Lens & Nantes & 2022-04-30 \\
                \midrule
                79 & Lorient & Reims & 2022-05-01 \\
                80 & Monaco & Angers SCO & 2022-05-01 \\
                81 & Montpellier & Metz & 2022-05-01 \\
                \midrule
                82 & Rennes & Saint-Etienne & 2022-04-30 \\
                83 & Troyes & Lille & 2022-05-01 \\
                84 & Brest & Strasbourg & 2022-05-07 \\
                \midrule
                85 & Lorient & Olympique Marseille & 2022-05-08 \\
                86 & Bordeaux & Lorient & 2022-05-14 \\
                87 & Olympique Lyonnais & Nantes & 2022-05-14 \\
                \midrule
                88 & Saint-Etienne & Reims & 2022-05-14 \\
                89 & Strasbourg & Clermont & 2022-05-14 \\
                90 & Troyes & Lens & 2022-05-14 \\
                \midrule
                91 & Angers SCO & Montpellier & 2022-05-21 \\
                92 & Brest & Bordeaux & 2022-05-21 \\
                93 & Clermont & Olympique Lyonnais & 2022-05-21 \\
                \midrule
                94 & Lens & Monaco & 2022-05-21 \\
                95 & Lille & Rennes & 2022-05-21 \\
                96 & Lorient & Troyes & 2022-05-21 \\
                \midrule
                97 & Olympique Marseille & Strasbourg & 2022-05-21 \\
                98 & Nantes & Saint-Etienne & 2022-05-21 \\
                99 & PSG & Metz & 2022-05-21 \\
                100 & Reims & Nice & 2022-05-21 \\
                \bottomrule
                \bottomrule
            \end{tabular}
        }
    \end{subtable}
    \caption{List of $100$ games used in this study. \textbf{Left}: Games 1-33. \textbf{Center}: Games 34-66. \textbf{Right}: Games 67-100.}
    \label{tab:games_combined}
\end{table}

\end{document}